\documentclass[11pt]{article}
\pdfoutput=1 
\usepackage[utf8x]{inputenc}
\usepackage[english]{babel}
\usepackage{graphicx}           
\usepackage{url}
\usepackage{eurosym}
\usepackage{makeidx}
\usepackage[plain]{algorithm}
\usepackage{algorithmic} 
\usepackage{color} 
\usepackage[margin=3.3cm]{geometry}
\usepackage{setspace}
\usepackage{datetime}
\usepackage{pifont}
\usepackage{tikz}
\usepackage{ulem}
\usepackage{natbib}
\usepackage{hyperref}
\usepackage{amsmath,amsfonts,amssymb,latexsym,stmaryrd}

\newcommand{\X}{\mathcal{X}}

\newcommand{\dif}{\mathrm{d}}
\newcommand{\rmd}{\dif}

\newcommand{\eqdef}     {\stackrel{{\textrm{\rm\tiny def}}}{=}}

\newcommand{\Sin}{{\mathbf s}_{\textrm{\tiny\rm in}}}

\newcommand{\taudiv}{b}
\newcommand{\mut}{\gamma}
\newcommand{\densiteIDE}{r}
\newcommand{\densiteIDEr}{m}

\newcommand{\Exp}{{\textrm{\rm Exp}}}
\newcommand{\tmax}{t_{\textrm{\tiny\rm max}}}
\newcommand{\rmax}{r_{\textrm{\tiny\rm max}}}

\newcommand{\mmax}{M}
\newcommand{\xdiv}{y}

\begin{document}

%%%%%%%%%%%%%%%%%%%%%%%%%%%%%%%%%%%%%%%%%%%%%%%%%%%%%%%%%%%%%%%%%%%%%%
\title{A numerical approach to determine mutant invasion fitness and evolutionary singular strategies}

\author{Coralie Fritsch$^{1,2,3}$ \and Fabien Campillo$^{4}$ \and Otso Ovaskainen$^{5,6}$}

\footnotetext[1]{CMAP, \'Ecole Polytechnique, UMR CNRS 7641, route de Saclay, Palaiseau Cedex, F-91128, France}

\footnotetext[2]{Universit\'e de Lorraine, Institut Elie Cartan de Lorraine,
    UMR CNRS 7502, Vand\oe uvre-l\`es-Nancy, F-54506, France}

\footnotetext[3]{Inria, TOSCA, Villers-l\`es-Nancy, F-54600, France}

\footnotetext[4]{Inria, LEMON, Montpellier, F-34095, France}

\footnotetext[5]{Department of Biosciences, FI-00014 University of Helsinki, Finland}

\footnotetext[6]{Centre for Biodiversity Dynamics, Department of Biology, Norwegian University of Science and Technology, N-7491 Trondheim, Norway \protect \\ 
				E-mail: {coralie.fritsch@inria.fr}, {fabien.campillo@inria.fr},
						 {otso.ovaskainen@helsinki.fi},
				 }
%%%%%%%%%%%%%%%%%%%%%%%%%%%%%%%%%%%%%%%%%%%%%%%%%%%%%%%%%%%%%%%%%%%%%%

%%%%%%%%%%%%%%%%%%%%%%%%%%%%%%%%%%%%%%%%%%%%%%%%%%%%%%%%%%%%%%%%%%%%%%
\maketitle
%%%%%%%%%%%%%%%%%%%%%%%%%%%%%%%%%%%%%%%%%%%%%%%%%%%%%%%%%%%%%%%%%%%%%%

\begin{abstract}
We propose a numerical approach to study the invasion fitness of a mutant and to determine evolutionary singular strategies in evolutionary structured models in which the competitive exclusion principle holds. 
Our approach is based on a dual representation, which consists of the modelling of the small size mutant population by a stochastic model and the computation of its corresponding deterministic model. The use of the deterministic model greatly facilitates the numerical determination of the feasibility of invasion as well as the convergence-stability of the evolutionary singular strategy.
Our approach combines standard adaptive dynamics with the link between the mutant survival criterion in the stochastic model and the sign of the eigenvalue in the corresponding deterministic model.
We present our method in the context of a mass-structured individual-based chemostat model.
We exploit a previously derived mathematical relationship between stochastic and deterministic representations of the mutant population in the chemostat model to derive a general numerical method for analyzing  the invasion fitness in the stochastic models.
Our method can be applied to the broad class of evolutionary models for which a link between the stochastic and deterministic invasion fitnesses can be established.


\paragraph{Mathematics Subject Classification (MSC2010):} 
92D15, 92D25, 35Q92, 45C05.
\end{abstract}

%%%%%%%%%%%%%%%%%%%%%%%%%%%%%%%%%%%%%%%%%%%%%%%%%%%%%%%%%%%%%%%%%%%%%%
%%%%%%%%%%%%%%%%%%%%%%%%%%%%%%%%%%%%%%%%%%%%%%%%%%%%%%%%%%%%%%%%%%%%%%
\section{Introduction}
%%%%%%%%%%%%%%%%%%%%%%%%%%%%%%%%%%%%%%%%%%%%%%%%%%%%%%%%%%%%%%%%%%%%%%
%%%%%%%%%%%%%%%%%%%%%%%%%%%%%%%%%%%%%%%%%%%%%%%%%%%%%%%%%%%%%%%%%%%%%%

Bacterial ecosystems are subject to mutations and natural selection. When a mutation occurs, a natural question is to determine the capability of the mutation to be fixed in the population. Adaptive dynamics theory proposes mathematical tools to tackle this question \citep{Metz1996a,dieckmann1996a,geritz1998a}. Among these tools,  invasion fitness is a selective value which allows to determine if a mutant population can invade a resident one. The definition of the invasion fitness depends on the model under consideration \citep{metz1992a,metz2008a}: usually for deterministic models, it is the asymptotic growth rate of the population,  for stochastic models we choose to define it as the survival probability of the mutant population, though we note that this is not a standard definition \citep{campillo2015a, campillo2016a}.

Here, we propose a general numerical approach to study the invasion capacity of a mutant population and to determine the evolutionary singular strategies when the competitive exclusion principle holds. The method is applied to a mass-structured chemostat model, for which we have obtained previous mathematical results which provide the necessary background information for developing our numerical approach. First developed by \cite{Monod1950a} and \cite{Novick1950a}, the chemostat is a culture method for maintaining a bacterial ecosystem in continuous growth. Adaptive dynamics in chemostat were studied for unstructured models in a deterministic context by \cite{Doebli2002,Mirrahimi2012} and in a stochastic context by \cite{champagnat2014a}.

The choice between stochastic and deterministic models usually depends on the context of the study: the latter for homogeneous large size populations, the former for small size populations where randomness cannot be neglected \citep{fritsch2015a, campillo2015b}. 
The numerical inference of adaptive dynamics for deterministic models is  straightforward compared to that of stochastic models. 
For example, for the stochastic mass-structured chemostat model, the invasion fitness is defined as the survival probability which is the solution of a functional equation. The mutant population can invade the resident one if and only if this survival probability is strictly positive. It is usually difficult to decide whether a numerical approximation of a survival probability is different from zero or not. For the corresponding deterministic model, the mutant population dynamics can be modeled as a population balance equation \citep{Fredrickson1967a, Ramkrishna1979a, Doumic2007a, Doumic2010a}, and the feasibility of invasion depends on the sign of the principal eigenvalue of the operator associated to this equation. The latter approach is more straightforward to apply, compared to evaluating the  positiveness of the survival probability.
\cite{campillo2015a, campillo2016a} established a mathematical link between the two invasion criteria for mass-structured growth-fragmentation-death models, that link considerably simplifies the numerical analysis of the stochastic version.

In this article, we focus on mutations which affect the division mechanism of a bacterial population; mutations can affect the mean proportion of the smallest daughter cell during the division and/or the minimal mass required for bacterial division.
We apply an evolutionary analysis to determine the best cytokinesis strategy that the population should adopt in terms of mass division. The optimal division strategy has been previously studied by \cite{Michel2005a, michel2006a}, in a deterministic context without interactions between bacteria (through the substrate in our model) and for strategies focusing only on the proportion of division.

In Section~\ref{sec.model}, we introduce the models which are considered for the numerical simulations as well as a model reduction approach, based on \cite{campillo2015b} and \cite{campillo2015a}.
In Section~\ref{sec.chemostat.model}, we introduce the mass-structured chemostat model where mutations affect the division parameters, namely the minimal mass for division and the mean proportion of the smallest daughter cell.
In Section~\ref{sec.to.reduced.model}, we reduce the chemostat model for the mutant population by assuming that the mutations are rare and the resident population is large. 
In Section~\ref{subec.reduced.model}, we introduce deterministic and stochastic representations for the reduced model of the mutant population. We present two definitions of the invasion fitness: the principal eigenvalue of a growth-fragmentation-death operator in the deterministic case, and the survival probability of the mutant population in the stochastic case. We also present the link between the invasion criteria derived from these two definitions, established by \cite{campillo2015a}.
Section~\ref{sec.num.methods} presents the numerical methods that we use for the simulations.
We present numerical tests in Section~\ref{sec.numeric}. In Section~\ref{sec.comparison}, we compare the different models, full chemostat model \textit{vs} reduced model and deterministic reduced model \textit{vs} stochastic reduced model, in order to numerically justify the model reduction of the mutant population and to present the difference between the deterministic and stochastic models.
In Section~\ref{sec.evol.c}, we study the evolution of the one-dimensional trait representing the mean proportion of the smallest daughter cell in the division mechanism. 
In Section~\ref{sec.evol.c.y}, we extend the approach presented in the previous section to  the case where both the mean proportion of the smallest daughter cell in the division mechanism and the minimal mass of division evolve simultaneously.
We conclude this article by a discussion in Section~\ref{ccl}.

%%%%%%%%%%%%%%%%%%%%%%%%%%%%%%%%%%%%%%%%%%%%%%%%%%%%%%%%%%%%%%%%%%%%%%
%%%%%%%%%%%%%%%%%%%%%%%%%%%%%%%%%%%%%%%%%%%%%%%%%%%%%%%%%%%%%%%%%%%%%%
\section{The models}
\label{sec.model}
%%%%%%%%%%%%%%%%%%%%%%%%%%%%%%%%%%%%%%%%%%%%%%%%%%%%%%%%%%%%%%%%%%%%%%
%%%%%%%%%%%%%%%%%%%%%%%%%%%%%%%%%%%%%%%%%%%%%%%%%%%%%%%%%%%%%%%%%%%%%%

This section details a numerical approach, based on common adaptive dynamics methods as well as mathematical results of \cite{campillo2015b} and \cite{campillo2015a}, which allows one to analyze evolutionary dynamics in a relatively complex chemostat model.

\medskip

Our approach is based on the following models that we will detail in this section:
\begin{itemize}
\item \textbf{IBM: individual-based model of resident and mutant populations.} This is the full chemostat model in which both resident and mutant populations are described by a stochastic individual-based model.
\item \textbf{PDE: deterministic approximation of the IBM.} Under a rare mutation assumption, between mutation times, the populations follow a system of integro-differential equations. This model is useful when populations are large, which is however not the case for the mutant population at a mutation time. A challenge associated with this model is to determine when a mutation occurs as the mutations are supposed to be rare.
\item \textbf{r-IBM: reduced individual-based model for the mutant population.} The resident population is assumed to stay at its stationary state, whereas the mutant population is described by an individual-based model. This model is realistic if the mutations are sufficiently rare in order for the resident population to reach its stationary state before the mutation
and as long as the mutant population remains sufficiently small to have a neglectable effect on the stationary state of the resident population.
\item \textbf{r-PDE: reduced deterministic model for the mutant population.} This model is a deterministic approximation of the r-IBM model: the resident population is assumed to stay at its stationary state and the mutant population evolves according to an integro-differential equation. This model may appear to be unrealistic due to opposing assumptions: the deterministic approximation is valid in large populations while the reduced model (constant resident population) is valid for a small mutant population. However, this model will prove to be very useful in the numerical study.
\end{itemize}

%%%%%%%%%%%%%%%%%%%%%%%%%%%%%%%%%%%%%%%%%%%%%%%%%%%%%%%%%%%%%%%%%%%%%%
\subsection{The chemostat model with mutations}
\label{sec.chemostat.model}
%%%%%%%%%%%%%%%%%%%%%%%%%%%%%%%%%%%%%%%%%%%%%%%%%%%%%%%%%%%%%%%%%%%%%%

We are interested to study numerically the evolution of a mass-structured population in a chemostat. We suppose that individuals grow by consuming a resource and divide after reaching a sufficiently large size. 
Individuals are also removed from the system due to the output flow of the chemostat. We also assume that during the division, mutations can appear in a gene that affects the minimal mass for division $\xdiv$ and/or the mean proportion $c$ of the smallest daughter bacteria after division (that is the expectation of the ratio of the size of the smaller of the two daughter cells over the size of the mother cell before division). A population will then be characterized by a trait $\xi=(\xdiv,c)\in\Xi$ which will evolve through mutations.

The model was originally introduced by \cite{fritschThesis} and \cite{campillo2015a}. In this model, each individual is characterized by its trait $\xi=(\xdiv,c)$ and its mass $x$. The model consists of the following stochastic events.
\begin{enumerate}
\item \textbf{Each individual divides} at rate $\taudiv(\xdiv,x)$ into two individuals with respective masses $\alpha \, x$ and $(1-\alpha)\,x$, where the proportion $\alpha$ is distributed according to a distribution $Q(c, \dif \alpha)=q(c, \alpha)\,\dif \alpha$, and
\begin{itemize}
\item with probability $\mut \in [0,1]$, the daughter cell with mass $\alpha\,x$ is a mutant, with trait $\xi+h \in \Xi$, where $h$ is distributed according to a distribution $\kappa(\xi, h)\,\dif h$ and the daughter cell with mass $(1-\alpha)\,x$ inherits the trait $\xi$ of its mother.

\item with probability $1-\mut$, the two daughter cells have the same trait $\xi$ as the mother cell.
\end{itemize}
\begin{center}
\includegraphics[width=6cm]{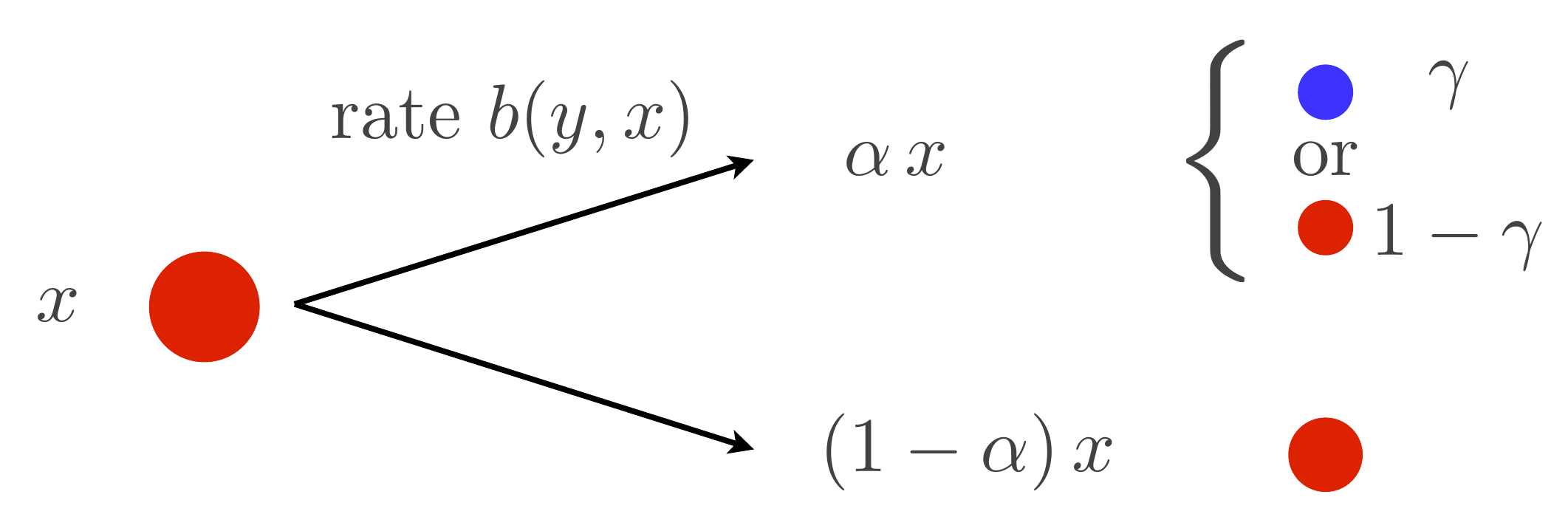}
\end{center}

\item \textbf{Each individual is removed} from the chemostat at rate $D$, due to the outflow, where $D$ is the dilution rate of the chemostat.

\item  \textbf{Each individual grows} at rate $g$, i.e.:
\begin{align}
  \frac{\rmd}{\rmd t} x_t 
  = 
  g(S_t, x_t)\,,
\end{align}
where $S_t$ represents the substrate concentration at time $t$ which evolves according~to
\begin{align}
\label{da.eq.substrat}
  \frac{\rmd}{\rmd t} S_t
  =
  D\,(\Sin-S_t) - \frac{k}{V} \, 
						\sum_{i=1}^{N_t} g(S_t,x_t^i)
\end{align}
where $\Sin$ is the substrate input concentration, $k$ is a stoichiometric coefficient, $V$ is the volume of the vessel and $x_t^1,\dots,\,x_t^{N_t}$ are the masses of the $N_t$ individuals in the chemostat at time $t$.
\end{enumerate}

\medskip

The model described above is a stochastic individual-based model (IBM) and a Markov process. To convert this model into a deterministic representation, we consider the above transition rates as the time-derivatives of partial differential equation (PDE).
What is however challenging is to derive a description of the mutation times as in a purely deterministic PDE model they cannot be modeled as random events.

The PDE approximation can be expected to be valid for large size populations, but a poor approximation for small sized ones, which is the very situation of the mutant population shortly after a mutation.

\bigskip

For the sake of simplicity we suppose that the growth function is a modified Gompertz one, where the growth rate depends on the substrate concentration and follows a Monod law at the level of the individual, namely:
\begin{align}
\label{eq.g}
  g(S,x)
  \eqdef
  r(S)\, \tilde g(x)
\end{align} 
with
\begin{align}
\label{eq.tile.g}
\tilde g(x)
	=
		 C_{a,d}\,\left[\log\left(\frac{\mmax}{x}\right)\right]^a\,x^d\,,
\end{align}
where $\mmax$ is the maximal size of the bacteria. The constant $C_{a,d}$ is defined by $C_{a,d} \eqdef \textstyle \frac \mmax e \, \left(\frac da \right)^a \, \frac {e^a}{\mmax^d}$ in order to keep the same maximal growth rate for the different values of $a$ and $d$. The function $r(S)$ is defined by
\begin{align}
\label{monod}
  r(S)
  \eqdef
  r_{\max} \, \frac{S}{K_r + S}
\end{align}
where $r_{\max}$ is the maximal growth rate and $K_r$ is the half velocity constant. 

\begin{figure}
\begin{center}
\includegraphics[width=3.9cm]{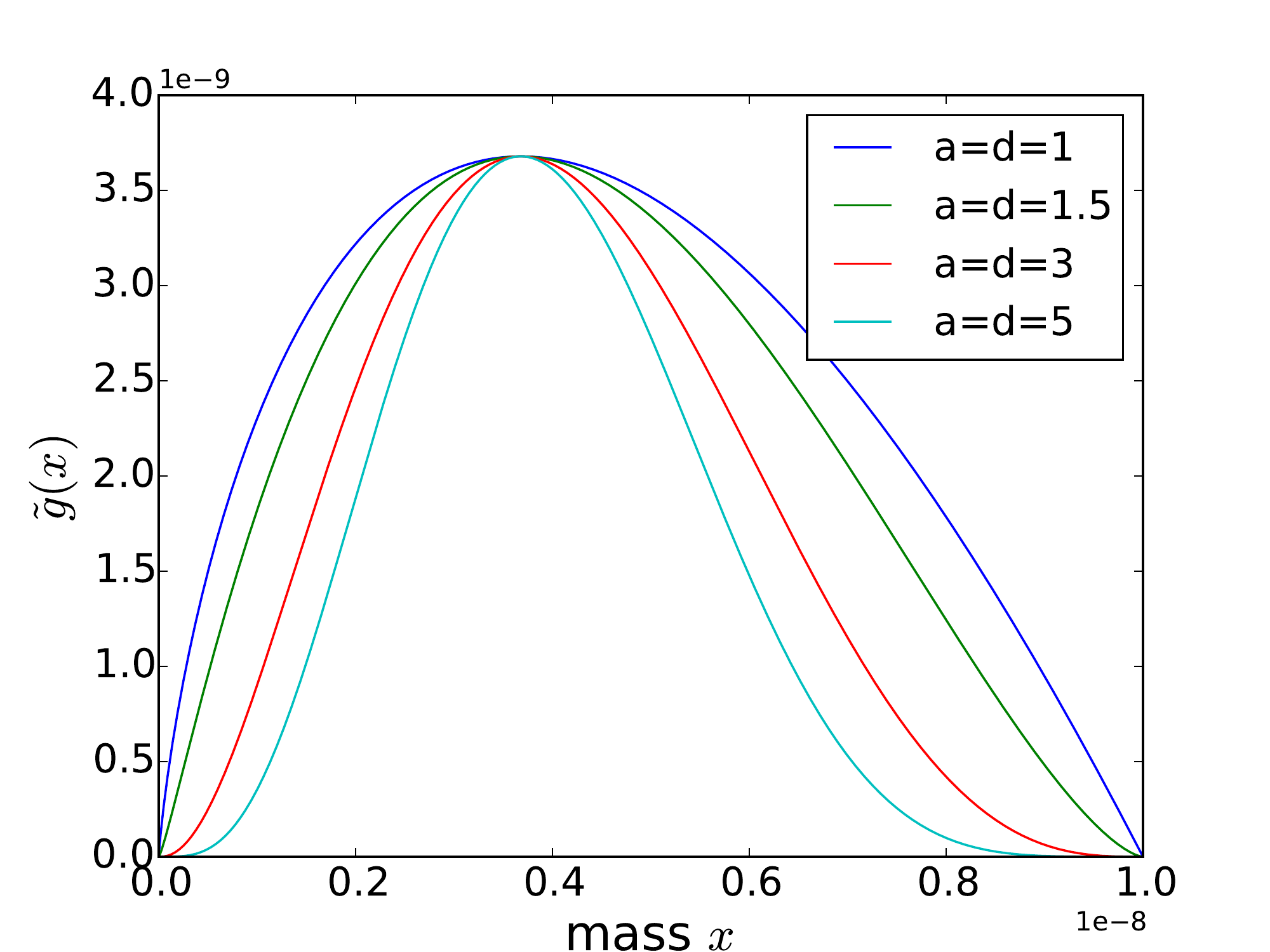}
\includegraphics[width=3.9cm]{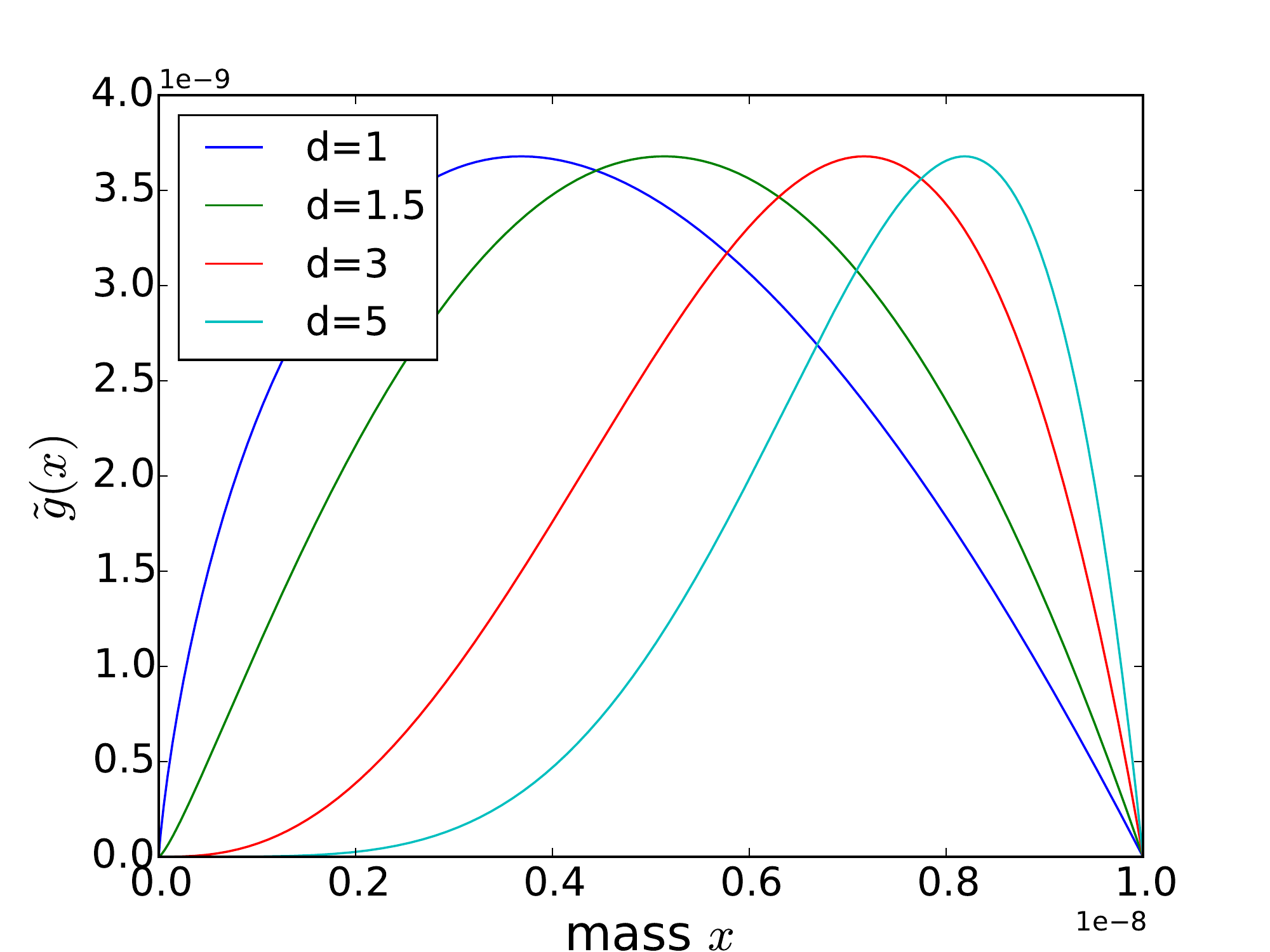}
\includegraphics[width=3.9cm]{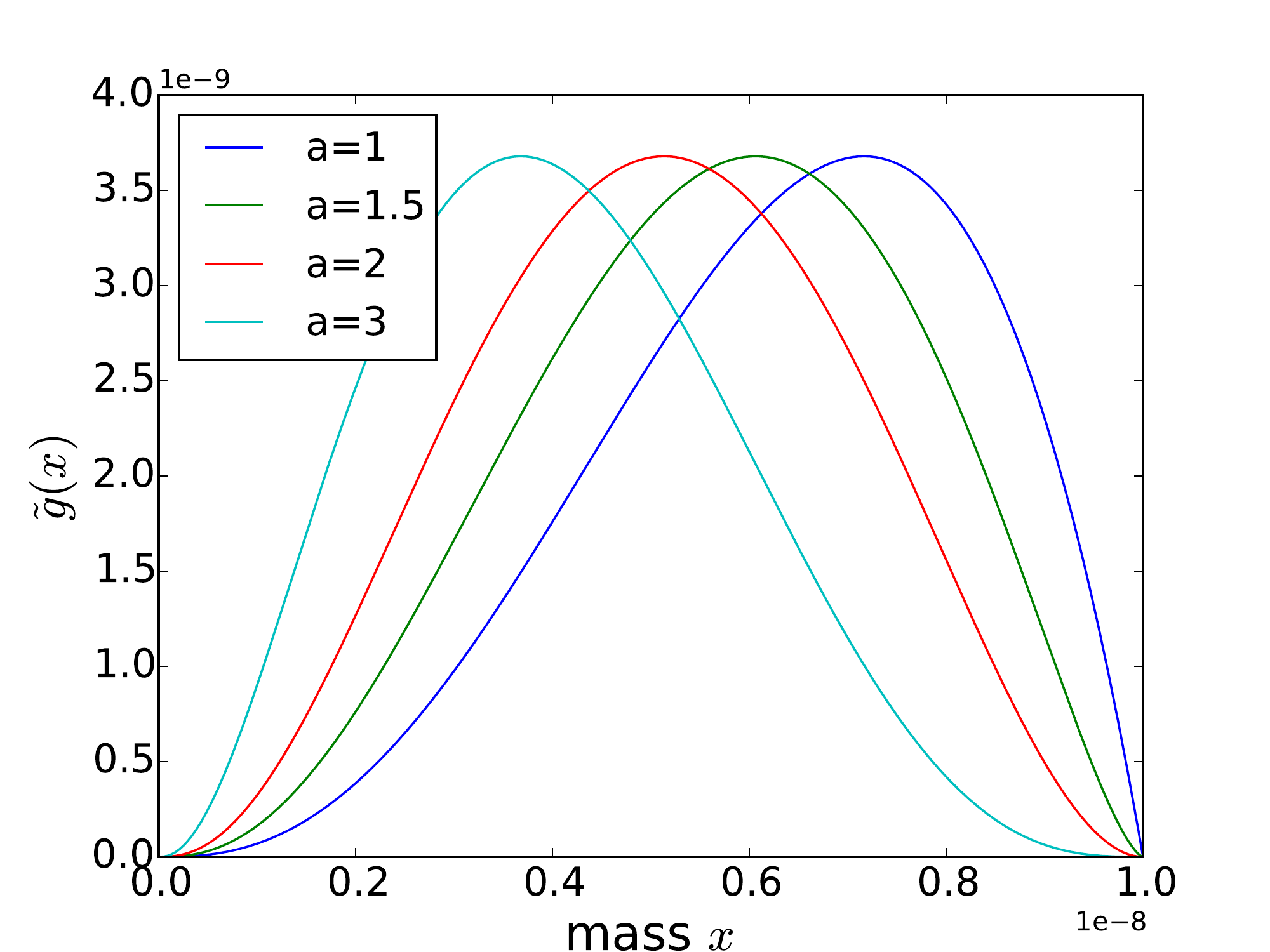}
\end{center}
\caption{$\blacktriangleright$ (Left) Growth function $\tilde g$ for $a=d=1$ (blue), $a=d=1.5$ (green), $a=d=3$ (red) and $a=d=5$ (cyan).
$\blacktriangleright$ (Center) Growth function $\tilde g$ for $a=1$ and $d=1$ (blue), $d=1.5$ (green), $d=3$ (red) and $d=5$ (cyan)
$\blacktriangleright$ (Right) Growth function $\tilde g$ for $d=3$ and $a=1$ (blue), $a=1.5$ (green), $a=2$ (red) and  $a=3$ (cyan).}
\label{fig.tilde.g}
\end{figure}

In \eqref{eq.tile.g}, the bigger the constant $d$ is, the more the growth of small individuals is penalized. The bigger the constant $a$ is, the more the growth of big individuals is penalized. The maximum growth speed is given by $r_{\max}\,\mmax/e$ and is reached for $x=\mmax\,e^{-a/d}$ when $S \to \infty$. Examples of such functions are given in Figure~\ref{fig.tilde.g}. For $a=d=1$, the function $\tilde g$ is the standard Gompertz function.

%---------------------------------------------
%---------------------------------------------

\bigskip

We assume that the division cannot occur below a minimum mass $\xdiv$. For individuals exceeding this size the division rate is independent of mass:
\begin{align*}
  \taudiv(\xdiv,x) = \bar\taudiv \,1_{\{x\geq \xdiv\}}\,.
\end{align*}

\begin{figure}[h]
\begin{center}
\includegraphics[width=6cm]{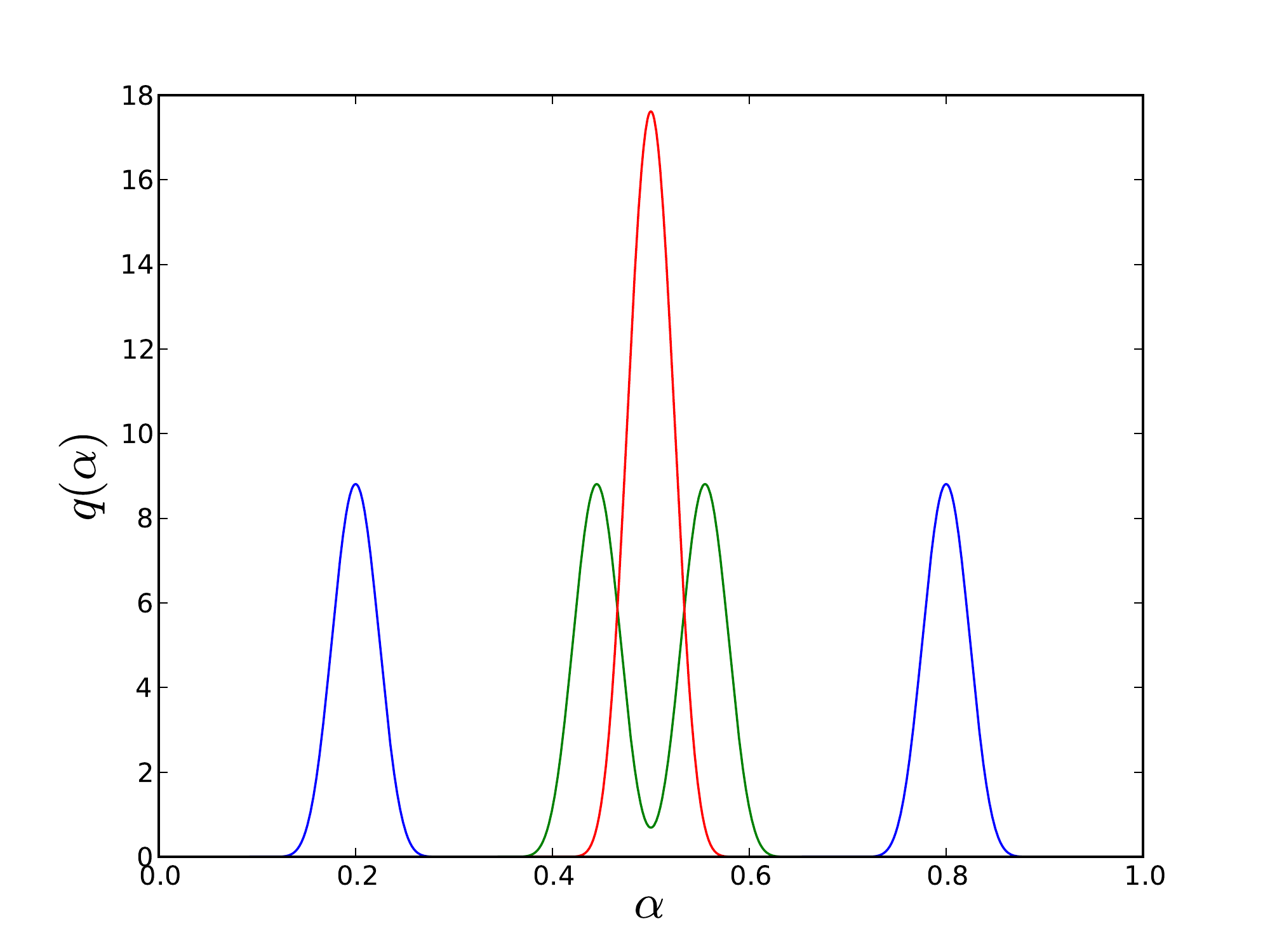}
\end{center}
\caption{Division kernels for $c=0.2$ (blue), $c=0.445$ (green) and $c=0.5$ (red). Other parameters are $l=0.1$ and $\theta=10$. }
\label{fig.kernel_q}
\end{figure}

The kernel $q$ giving the  proportion $\alpha$ of mass allocated to one of the two offspring is:
\begin{align*}
  q(c,\alpha)
  &=
  \frac{1}{C_{\theta,c, l}}\,
  \bigl\{ (\alpha-(c-l))^{\theta-1} \, (c+l-\alpha)^{\theta-1} 
		\, 1_{\{\alpha \in [c-l, c+l]\}}
  \\[-0.7em]
  &\qquad\qquad
				+ (\alpha-(1-c-l))^{\theta-1} \, (1-c+l-\alpha)^{\theta-1}
				\, 1_{\{\alpha \in [1-c-l, 1-c+l]\}}
  \bigr\} 	 
\end{align*}
where $C_{\theta,c, l}$ is a normalizing constant. The constants $l$ and $\theta$ give the variance of the division proportion and $0\leq c\leq 0.5$ represents the mean proportion of the smaller offspring. Examples of such kernels are drawn in Figure~\ref{fig.kernel_q}.
If $c=0.5$ then the division processes is a noisy symmetric division, if not, the division is asymmetric.

%---------------------------------------------
%---------------------------------------------

\subsection{Deriving the reduced stochastic model (r-IBM) from the full chemostat model (IBM)}
\label{sec.to.reduced.model}

Assume that the initial population is monomorphic with trait $\xi_R=(\xdiv_R,c_R)$ and that this population is large. Then, until the first mutation time, according to \cite{campillo2015b}, this monomorphic population can be approximated by the following deterministic model (PDE before the first mutation time):
\begin{align} 
\label{eq.eid.substrat}
	&
	\frac{\rmd}{\rmd t} S_t  = 
	D\,(\Sin-S_t)-\frac kV \int_0^{\mmax} g(S_t,x)\,
	\densiteIDE_t(x)\,\dif x\,,
\\
\nonumber
	&
	\frac{\partial}{\partial t} \densiteIDE_t(x)
	+\frac{\partial}{\partial x} \bigl( g(S_t,x)\,\densiteIDE_t(x)\bigr)
	+ \bigl(\taudiv(\xdiv_R,x)+D \bigr)\,\densiteIDE_t(x)
\\
\label{eq.eid.pop}
  	&\qquad\qquad\qquad\qquad\qquad\qquad\qquad
	=  
	2\,\int_x^{\mmax}
		\frac{\taudiv(\xdiv_R,z)}{z} \, 
		q\left(c_R,\frac{x}{z} \right) \,
		\densiteIDE_t(z)\,\dif z \,,
\end{align}
where $\densiteIDE_t$ represents the mass density of the population at time $t$.  

Moreover, if the mutations are rare enough, we can assume that the chemostat reaches a neighborhood of its stationary state $(S^*_{\xi_R}, r^*_{\xi_R})$ before a mutation occurrence, where $S^*_{\xi_R}$ and $r^*_{\xi_R}$ are the substrate concentration and the mass density at the equilibrium when the chemostat evolves with only one population with trait $\xi_R$ \citep{campillo2015a}.

The population with trait $\xi_R$ will be named the resident population. Just after a mutation, the size of the mutant population, with trait $\xi_M$, is negligible with respect to the size of the resident one. The effect of the mutant population on the stationary state $(S^*_{\xi_R}, r^*_{\xi_R})$ is then negligible (see \cite{geritz1998a}). As the mutations are assumed to be rare, secondary mutations can be neglected during the invasion attempt of a primary mutant. Then the mutant population can, as long as it is small compared to the resident population, be described as evolving in a constant environment given by the equilibrium $(S^*_{\xi_R}, r^*_{\xi_R})$ and affected by the division mechanism (without mutation), the death mechanism and the growth at speed $g(S^*_{\xi_R},.)$ (see the reduced model described in Section~\ref{subec.reduced.model}).

Two cases are possible when a mutation occurs:
\begin{itemize}
\item \textbf{1st case :} The resident population is ``better'' than the mutant population, in the sense that the resident population has a better fitness/growth that the mutant one. Hence the mutant population goes extinct.

\item \textbf{2nd case :} The mutant population is ``better'' than the resident population, hence the mutant population may invade the resident one. In this case, there are two possibilities:
\begin{itemize}
\item the mutant population goes extinct even though the equilibrium $(S^*_{\xi_R}, r^*_{\xi_R})$ is in favor of the invasion of the mutant population;
\item the mutant population invades the resident population and, as soon as the two populations are large enough, the dynamics of the chemostat can then be approached by the following system (PDE after the mutation time):
\begin{align} 
\label{eq.eid.substrat.2pop}
	&
	\frac{\rmd}{\rmd t} S_t  = 
	D\,(\Sin-S_t)-\frac kV \int_0^{\mmax} g(S_t,x)\,
	(\densiteIDE_{R,t}(x)+\densiteIDE_{M,t}(x))\,\dif x\,,
\\
\nonumber
	&
	\frac{\partial}{\partial t} \densiteIDE_{R,t}(x)
	+\frac{\partial}{\partial x} \bigl( g(S_t,x)\,\densiteIDE_{R,t}(x)\bigr)
	+ \bigl(\taudiv(\xdiv_R,x)+D \bigr)\,\densiteIDE_{R,t}(x)
\\
\label{eq.eid.pop.res}
  	&\qquad\qquad\qquad\qquad\qquad\qquad\qquad
	=  
	2\,\int_x^{\mmax}
		\frac{\taudiv(\xdiv_R,z)}{z} \, 
		q\left(c_R,\frac{x}{z} \right) \,
		\densiteIDE_{R,t}(z)\,\dif z \,,
		\\
\nonumber
	&
	\frac{\partial}{\partial t} \densiteIDE_{M,t}(x)
	+\frac{\partial}{\partial x} \bigl( g(S_t,x)\,\densiteIDE_{M,t}(x)\bigr)
	+ \bigl(\taudiv(\xdiv_M,x)+D \bigr)\,\densiteIDE_{M,t}(x)
\\
\label{eq.eid.pop.res.mut}
  	&\qquad\qquad\qquad\qquad\qquad\qquad\qquad
	=  
	2\,\int_x^{\mmax}
		\frac{\taudiv(\xdiv_M,z)}{z} \, 
		q\left(c_M,\frac{x}{z} \right) \,
		\densiteIDE_{M,t}(z)\,\dif z \,,
\end{align}
where $\densiteIDE_{R,t}$ and $\densiteIDE_{M,t}$ are respectively the densities of the resident and the mutant populations at time $t$. Thus the resident population goes  extinct and the substrate/mutant population pair reaches a neighborhood of its stationary state. The mutant population then becomes the resident population until the next mutation.
\end{itemize}  

\end{itemize}

%%%%%%%%%%%%%%%%%%%%%%%%%%%%%%%%%%%%%%%%%%%%%%%%%%%%%%%%%%%%%%%%%%%%%%
\subsubsection{Description of the reduced growth-fragmentation-death model (r-IBM and r-PDE)}
\label{subec.reduced.model}
%%%%%%%%%%%%%%%%%%%%%%%%%%%%%%%%%%%%%%%%%%%%%%%%%%%%%%%%%%%%%%%%%%%%%%

We made the following assumptions in the previous section: monomorphic initial/resident population with trait $\xi_R$, large resident population and rare mutations.
Therefore, a mutant population, with trait $\xi=(\xdiv,c)$, can be represented, as long it is in a small population size, by the following growth-fragmentation-death model: one individual with mass $x$,
\begin{enumerate}
\item \textbf{divides} at rate $\taudiv(\xdiv,x)$, into two individuals with masses $\alpha \, x$ and $(1-\alpha)\,x$, where the proportion $\alpha$ is distributed according to the distribution $Q(c, \dif \alpha)=q(c, \alpha)\,\dif \alpha$;
\begin{center}
\includegraphics[width=5cm]{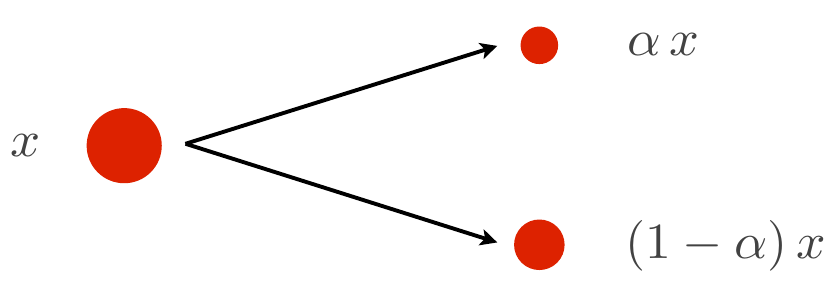}
\end{center}

\item \textbf{is withdrawn} from the chemostat at rate $D$;

\item  \textbf{grows} at speed $g(S^*_{\xi_R}, .)$:
\begin{align}
\label{eq.g.reduced}
  \frac{\rmd}{\rmd t} x_t 
  = 
  g(S^*_{\xi_R}, x_t)
  = r(S^*_{\xi_R})\, \tilde g(x_t)\,.
\end{align}
\end{enumerate}

\bigskip

We will consider two representations of this growth-fragmentation-death model. 
The first is a stochastic individual-based model (r-IBM) and the second   is a deterministic integro-differential model (r-PDE).
When a mutant population appears in the chemostat, it is in small population size, and thus submitted to a non-neglectable randomness that can be captured by a stochastic model only. In particular, the probability that the mutant goes extinct is non-neglectable although the environment is in favor of its invasion. However, the stochastic representation is more complex to study numerically and analytically than the deterministic integro-differential model. The mathematical link between these two representations (see Section \ref{links.between.fitness}), established by \cite{campillo2015a}, allows one to use the deterministic model to study invasion criteria for the stochastic model.

\subsubsection{Invasion fitness for the r-PDE}

The deterministic integro-differential model is described by the following equation:
\begin{align} 
\label{da.eid.reduit}
\nonumber
&	\frac{\partial}{\partial t} \densiteIDEr_t(x)
	+\frac{\partial}{\partial x} \bigl(g(S^*_{\xi_R}, x)\,\densiteIDEr_t(x)\bigr)
	+ \bigl(\taudiv(\xdiv,x)+D \bigr)\,\densiteIDEr_t(x)
\\
	& \qquad \qquad\qquad \qquad \qquad \qquad=  
	2\,\int_0^{\mmax} 
		\frac{\taudiv(\xdiv,z)}{z}\, 
		q\left(c,\frac{x}{z}\right)\, 
		\densiteIDEr_t(z)\,\dif z
\end{align}
where $\densiteIDEr_t$ represents the mass density of the population at time $t$.
The invasion fitness is defined as the exponential growth rate of the population and it corresponds to the eigenvalue $\Lambda(S^*_{\xi_R},\xi)$ of the following eigenproblem (see for example \cite{Doumic2007a, perthame2007a, campillo2015a}):
\begin{subequations}
\label{eq.eigenproblem}
\begin{align}
\nonumber  &
  \frac{\partial}{\partial x} \bigl(g(S^*_{\xi_R}, x)\, N_{\xi_R}(\xi, x)\bigr)
  +
  (\taudiv(\xdiv,x)+D+\Lambda(S^*_{\xi_R},\xi))\,N_{\xi_R}(\xi,x)
\\\label{eq.eigenproblem.1}
  &\qquad\qquad\qquad\qquad\qquad\qquad
-2\,\int_\X \frac{\taudiv(\xdiv,z)}{z}\, q\left(c,\frac{x}{z}\right)
	\, N_{\xi_R}(\xi,z)\,\dif z = 0
\end{align}
with conditions:
\begin{align}
\label{eq.eigenproblem.2}
   g(S^*_{\xi_R},0)\, N_{\xi_R}(\xi,0) &= 0\,,
   & 
   D+\Lambda(S^*_{\xi_R},\xi) &>0\,,
   & 
   \int_\X N_{\xi_R}(\xi,x)\,\dif x &= 1
\end{align}
\end{subequations}
where $N_{\xi_R}(\xi, x)$ is the associated eigenvector. The sign of the eigenvalue $\Lambda(S^*_{\xi_R},\xi)$ determines if the mutant population can invade or not the resident one:
\begin{itemize}
\item[\textbullet] if $\Lambda(S^*_{\xi_R},\xi)>0$, then the mutant population can invade the resident one;
\item[\textbullet] if $\Lambda(S^*_{\xi_R},\xi)\leq 0$ then the mutant population cannot invade the resident one.
\end{itemize}

In the rest of the paper, we will consider situations where the equilibria $(S^*_{\xi_R}, r^*_{\xi_R})$ are non-trivial, that is, the monomorphic population with trait $\xi_R$ does not go extinct in the chemostat model without mutation. Hence, $\Lambda(S^*_{\xi_R},\xi_R)=0$ as the resident population is assumed to be at equilibrium.

%%%%%%%%%%%%%%%%%%%%%%%%%%%%%%%%%%%%%%%%%%%%%%%%%%%%%%%%%%%%%%%%%%%%%%
\subsubsection{Invasion fitness for the r-IBM}
%%%%%%%%%%%%%%%%%%%%%%%%%%%%%%%%%%%%%%%%%%%%%%%%%%%%%%%%%%%%%%%%%%%%%%

For the stochastic individual-based model, described in \cite{campillo2015a}, the invasion fitness can be defined as the survival (or invasion) probability of the mutant population described by the reduced model. 

\cite{campillo2015a} proved that this survival probability $p_{S^*_{\xi_R}}^{\xi}(x)$ depends on the mass $x$ of the initial mutant individual and is a functional solution of
\begin{align}
\label{da.eq.proba.survie}
  p_{S^*_{\xi_R}}^{\xi}(x)
  &=
  \int_0^\infty \taudiv(\xdiv,A_t(x)) \, e^{-\int_0^t \taudiv(\xdiv,A_u(x)) 
      \,\dif u -D\,t}\,
\\
\nonumber
	& \qquad\qquad
	\int_0^1 q(c,\alpha)\,
		\Big[
			p_{S^*_{\xi_R}}^{\xi}\big(\alpha \, A_t(x)\big)
	     	+ p_{S^*_{\xi_R}}^{\xi}\big((1-\alpha) \, A_t(x)\big)
\\
\nonumber
	& \qquad\qquad\qquad\qquad\qquad
			- p_{S^*_{\xi_R}}^{\xi}\big(\alpha \, A_t(x)\big) \, p\big((1-\alpha) \, 
				A_t(x)\big)   	
	     \Big]\, 
	     \dif \alpha\, \dif t
\end{align}
where $A_t$ is the flow associated to the growth at speed $g$, i.e. for any $t>0$ and $x\in[0,\mmax]$:
\begin{align}
\label{def.flot}
  \frac{\dif}{\dif t}A_t(x) = g\left(S^*_{\xi_R},A_t(x)\right)\,,
  \qquad A_0(x)=x\,.
\end{align}
If $p_{S^*_{\xi_R}}^{\xi}(x)>0$, then the mutant population can invade the resident one. If $p_{S^*_{\xi_R}}^{\xi}(x)=0$, then the mutant population goes extinct almost surely.

Unfortunately, \eqref{da.eq.proba.survie} does not allow one to obtain an explicit representation of the invasion fitness.

%%%%%%%%%%%%%%%%%%%%%%%%%%%%%%%%%%%%%%%%%%%%%%%%%%%%%%%%%%%%%%%%%%%%%%
\subsubsection{Link between the invasion fitnesses of the r-PDE and the r-IBM}
\label{links.between.fitness}
%%%%%%%%%%%%%%%%%%%%%%%%%%%%%%%%%%%%%%%%%%%%%%%%%%%%%%%%%%%%%%%%%%%%%%

The eigenvalue $\Lambda(S^*_{\xi_R},\xi)$ of  problem~\eqref{eq.eigenproblem} is more straightforward to compute than a solution of~\eqref{da.eq.proba.survie}. Thus, making the link between these two invasion fitnesses is very useful in determining when a mutant population can invade a resident one for the stochastic model.
\cite{campillo2015a}  made this link in the following sense:
\begin{align*}
\Lambda(S^*_{\xi_R},\xi)>0 \quad \Longleftrightarrow \quad p_{S^*_{\xi_R}}^{\xi}(x)>0,\,
\text{for any } x\in(0,\mmax)\,. 
\end{align*}
Thus, the invasion criteria in the stochastic and deterministic models are in this sense equivalent.
However, the eigenvalue $\Lambda(S^*_{\xi_R},\xi)$ does not depend on the initial mass $x$ whereas the invasion probability $p_{S^*_{\xi_R}}^{\xi}(x)$ does depend on it.

Contrary to the full chemostat model, in which individuals indirectly interact through the substrate consumption, there is no interaction between individuals in the reduced models. The stochastic reduced model is then a branching process. This is the key argument in making the link between the stochastic and the deterministic reduced models \citep{campillo2015a}.

%%%%%%%%%%%%%%%%%%%%%%%%%%%%%%%%%%%%%%%%%%%%%%%%%%%%%%%%%%%%%%%%%%%%%%
%%%%%%%%%%%%%%%%%%%%%%%%%%%%%%%%%%%%%%%%%%%%%%%%%%%%%%%%%%%%%%%%%%%%%%
\section{Numerical methods}
\label{sec.num.methods}
%%%%%%%%%%%%%%%%%%%%%%%%%%%%%%%%%%%%%%%%%%%%%%%%%%%%%%%%%%%%%%%%%%%%%%
%%%%%%%%%%%%%%%%%%%%%%%%%%%%%%%%%%%%%%%%%%%%%%%%%%%%%%%%%%%%%%%%%%%%%%

We now describe the numerical methods that we will use in Section~\ref{sec.numeric} where we examine numerically the links between the model variants.
For many of these methods, we make use of the algorithm and the associated \texttt{Python}-code both developed by \cite{fritsch2015a} to solve the system~\eqref{eq.eid.substrat}-\eqref{eq.eid.pop}. The algorithm is based on an explicit Euler scheme for the time integration coupled to an uncentered upwind difference scheme for the space integration.

\medskip

The PDE~\eqref{eq.eid.substrat.2pop}-\eqref{eq.eid.pop.res.mut} is simulated according to the scheme proposed in \cite{fritsch2015a} extended to the case of two populations.

\medskip

For the simulation of the r-IBM of a mutant population with trait $\xi=(c,y)$, we solved the system~\eqref{eq.eid.substrat}-\eqref{eq.eid.pop} for the resident population numerically until $t$ is sufficiently large to reach an equilibrium $(S^*_{\xi_R}, r^*_{\xi_R})$; we then simulated the mutant population by the standard Gillespie algorithm:
\begin{algorithmic}
\STATE compute $\nu_0=\{x^1_{0}\}$
 \COMMENT{initial mutant individual}
\STATE $t\leftarrow 0$
\STATE $N\leftarrow 1$ \COMMENT{initial mutant population size}
\WHILE {$t\leq \tmax$}
  \STATE $\tau \leftarrow (\bar b+D)\,N$
  \STATE $\Delta t \sim \Exp(\tau)$
  \STATE compute $\nu_{t+\Delta t}$ integrating \eqref{eq.g.reduced}  over $[t,t+\Delta t]$ 
  		for each $x \in \{x^i_{t}\,;\,i=1,\dots,N\}$ 
  \STATE $t \leftarrow t+\Delta t$
  \STATE draw $x$ uniformly in $\{x^i_{t}\,;\,i=1,\dots,N\}$  
  \STATE $u\sim U[0,1]$ 
  \IF {$u\leq b(y,x)/(\bar b+D)$}
      \STATE $\alpha \sim q(c,.)$
      \STATE $\nu_{t} \leftarrow \left(\nu_{t}\setminus\{x\}\right)
      		\cup\{\alpha\,x, \, (1-\alpha)\,x\}$
      \COMMENT{division}
      \STATE $N\leftarrow N+1$
  \ELSIF{$u\leq (b(y,x)+D)/(\bar b+D)$}
      \STATE $\nu_{t} \leftarrow \nu_{t}\setminus\{x\}$ 
      \COMMENT{up-take}
      \STATE $N\leftarrow N-1$
  \ENDIF
\ENDWHILE
\end{algorithmic}

\medskip

To solve the r-PDE~\eqref{da.eid.reduit},
\begin{itemize}
\item[\textbullet] We solved the equations~\eqref{eq.eid.substrat}-\eqref{eq.eid.pop} for the resident population numerically until $t$ was sufficiently large so that the system had reached an equilibrium $(S^*_{\xi_R}, r^*_{\xi_R})$;
\item[\textbullet] next, we solved equation~\eqref{da.eid.reduit} numerically for $t$ sufficiently large so that the normalized density reached an equilibrium;
\item[\textbullet] finally, the eigenvalue $\Lambda(S^*_{\xi_R},\xi_M)$ is obtained as the exponential growth rate of the population. It corresponds to the value such that $\dif N_t/\dif t=\Lambda(S^*_{\xi_R},\xi_M)\,N_t$ where $N_t=\int_0^\mmax m_t(x)\,\dif x$ is the number of mutant individuals at time $t$ for $t$ large enough.
\end{itemize}

%%%%%%%%%%%%%%%%%%%%%%%%%%%%%%%%%%%%%%%%%%%%%%%%%%%%%%%%%%%%%%%%%%%%%%
%%%%%%%%%%%%%%%%%%%%%%%%%%%%%%%%%%%%%%%%%%%%%%%%%%%%%%%%%%%%%%%%%%%%%%
\section{Numerical results}
\label{sec.numeric}
%%%%%%%%%%%%%%%%%%%%%%%%%%%%%%%%%%%%%%%%%%%%%%%%%%%%%%%%%%%%%%%%%%%%%%
%%%%%%%%%%%%%%%%%%%%%%%%%%%%%%%%%%%%%%%%%%%%%%%%%%%%%%%%%%%%%%%%%%%%%%

We performed simulations with the parameters of Table~\ref{table.parametres.comp}.
Other values can produce similar behaviors, but of course there are values for which the model does not exhibit noteworthy behaviors, our aim is not to explore all the parameter space.

%-----------------------------------
\begin{table}[h]
\begin{center}
\begin{tabular}{|c|c|}
	\hline
    Parameters & Values \\
    \hline
    $\Sin$				&	10 mg/l \\
    $M$					&	$1.0\times10^{-8}$ mg \\	
    $\xdiv$				&	$0.5\times10^{-8}$ mg \\
	$\bar b$		&	20 h$^{-1}$\\
	$\theta$				&	10 \\
	$l$					&	0.1 \\
    $\rmax$				&	1 h$^{-1}$\\
    $K_r$				&	10 mg/l\\
    $k$					&	100\\
    $D$					& 	0.31 h$^{-1}$\\
    $V$					& 	0.1 l\\
    $a$					&	1.4\\
    %$d$					&	1.8\\
    \hline
\end{tabular}
\end{center}
\caption{Simulation parameters.}
\label{table.parametres.comp}
\end{table}
%-----------------------------------

\subsection{Comparison of the different models}
\label{sec.comparison}

\begin{figure}
\begin{center}
\includegraphics[width=6cm]{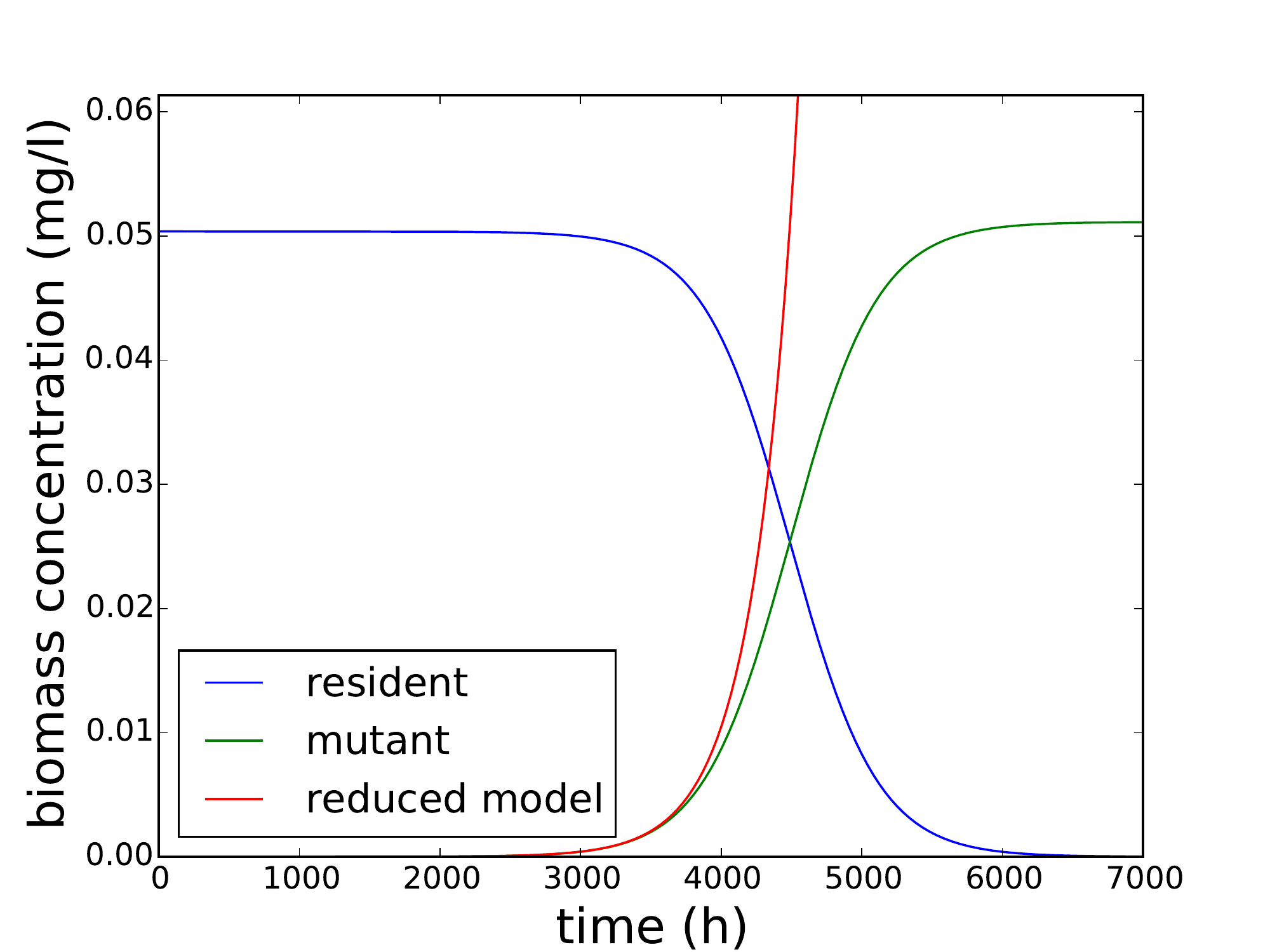}
\includegraphics[width=6cm]{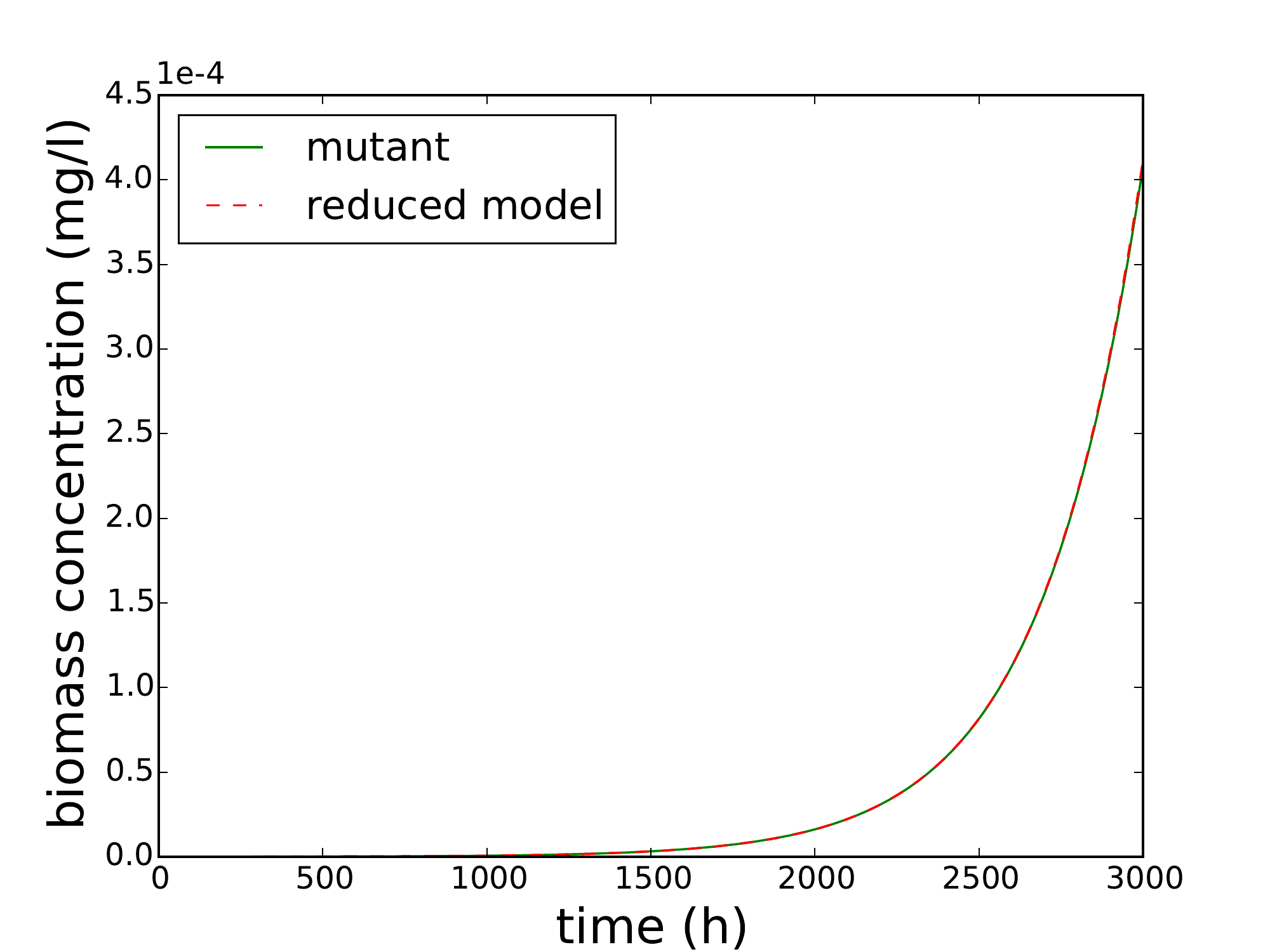}
\end{center}
\caption{Time evolution of the biomass concentrations of the resident population with $\xi_R=(0.0005,0.5)$ (blue) and the mutant population with $\xi_M=(0.0005,0.2)$ (green) for the PDE and of the mutant population with $\xi_M=(0.0005,0.2)$ (red) for the r-PDE, until time $t=7000$ h (left), with parameters of Table~\ref{table.parametres.comp} and $d=1.8$. The figure on the right is a zoom on the mutant biomass concentration evolutions for the two models until time $t=3000$ h.}
\label{fig.biomass.evolution}
\end{figure}

Figure~\ref{fig.biomass.evolution} describes the time evolution of the biomass concentrations of a resident population (blue) and of a mutant population (green) for the PDE~\eqref{eq.eid.pop.res}-\eqref{eq.eid.pop.res.mut} with initial condition $(S_0,r_{R,0})=(S_{\xi_R^*},S_{r_R^*})$ which is the equilibrium of the system~\eqref{eq.eid.substrat}-\eqref{eq.eid.pop}, as well as the time evolution of the biomass of the mutant population given by the r-PDE~\eqref{da.eid.reduit}, with parameters of Table~\ref{table.parametres.comp} and with $d=1.8$. As soon as the number of individuals of the mutant population is not neglectable (after $t=3000$ h) the time evolution of the biomass of the mutant population of the two models are different. Before this time, the two models behave very closely (Figure~\ref{fig.biomass.evolution}, right). This provides support for the use of the reduced model for the initial phase of the mutant invasion.

%---------------------------------------------
%---------------------------------------------

\bigskip

%---------------------------------------------
\begin{figure}
\begin{center}
\includegraphics[width=6cm]{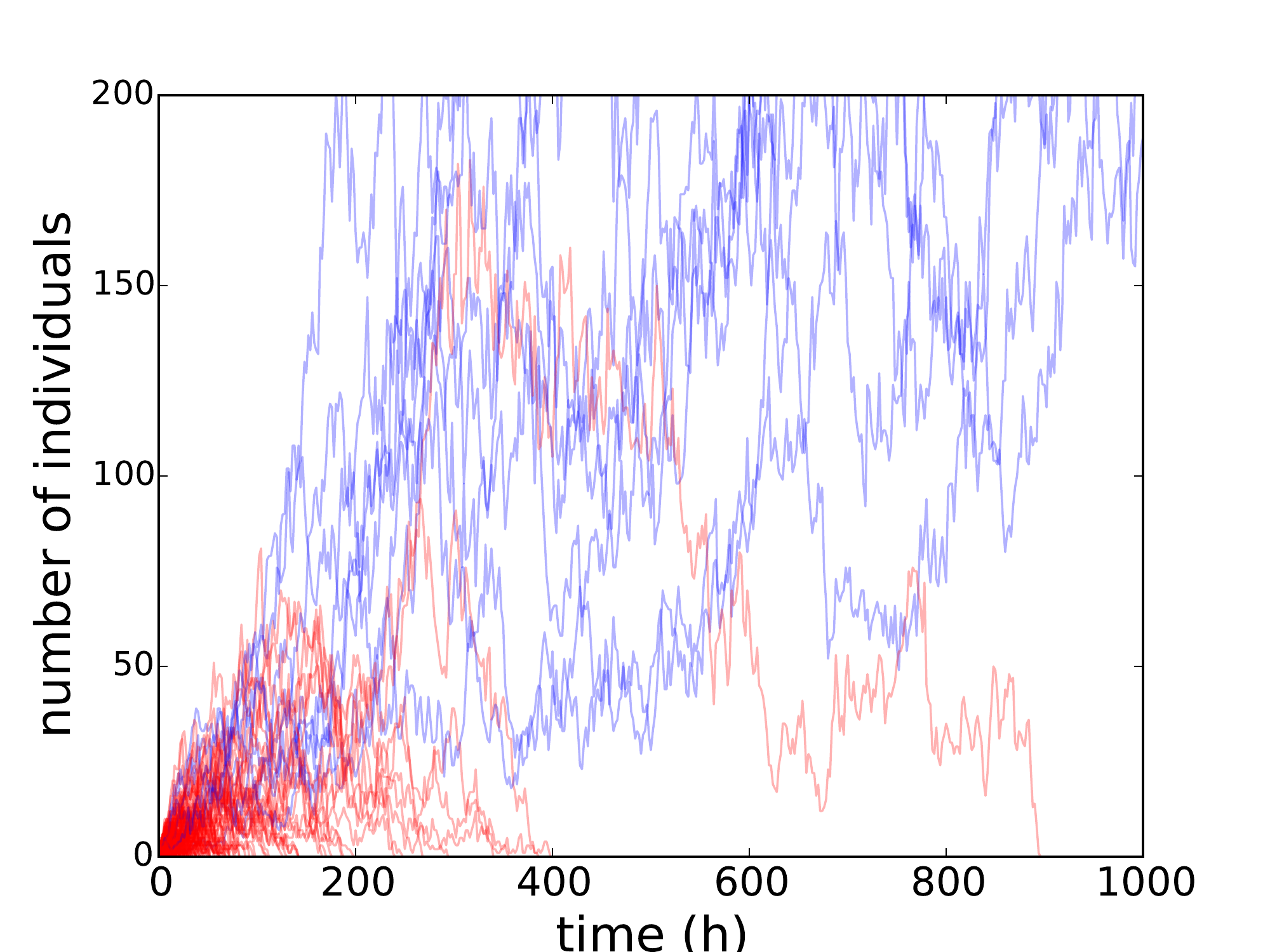}
\includegraphics[width=6cm]{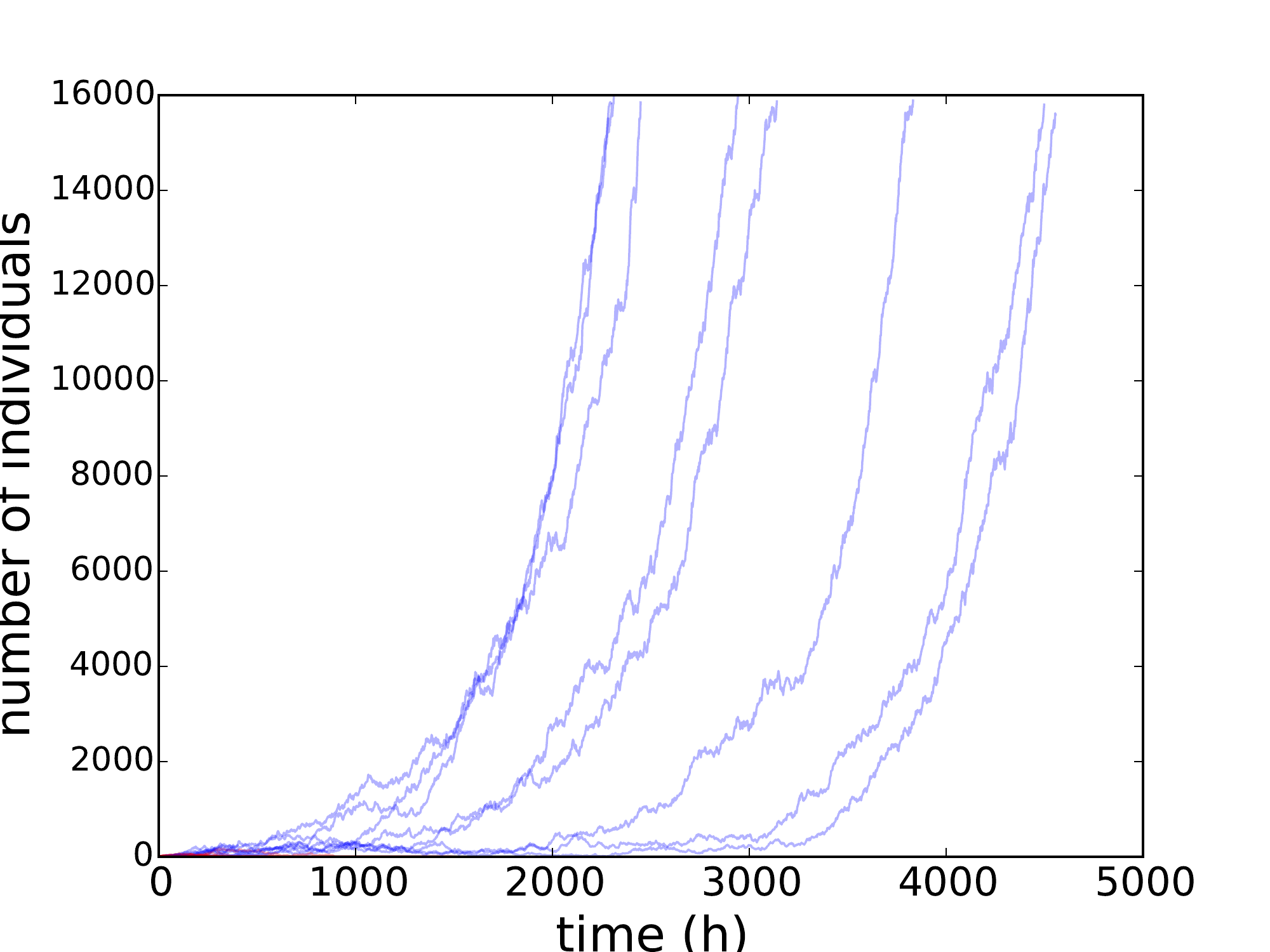}
\end{center}
\caption{Time evolution of the number of individuals of 500 independent simulations for the r-IBM until time $t=1000$ h (left) and $t=5000$ h (right).}
\label{fig.biomass.ibm.reducedModel}
\end{figure}
%---------------------------------------------

\begin{figure}
\begin{center}
\includegraphics[width=6cm]{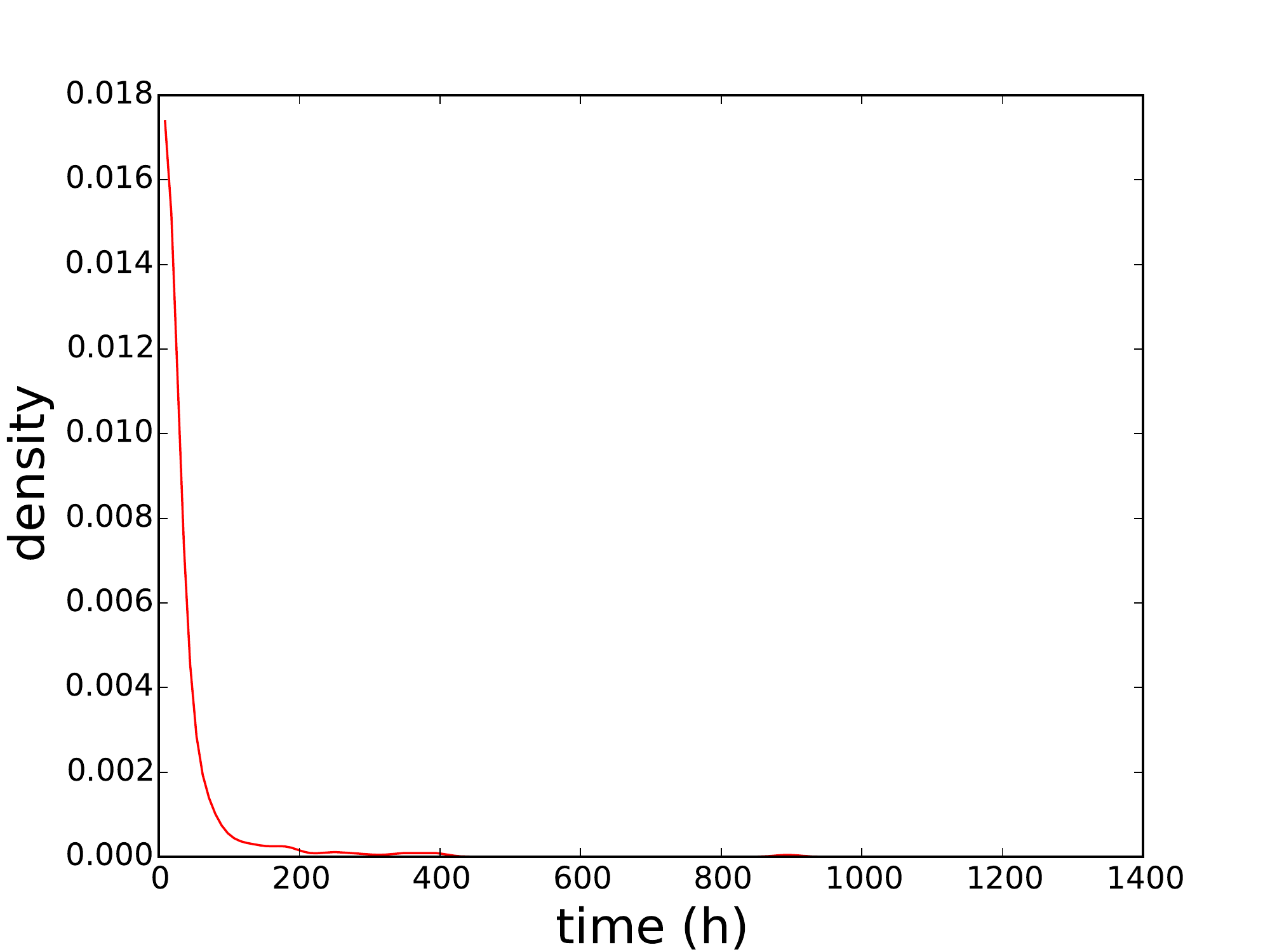}
\end{center}
\caption{Empirical law of the extinction time of the 492 extinct populations of the r-IBM.}
\label{fig.extinction.time}
\end{figure}

Figure~\ref{fig.biomass.ibm.reducedModel} represents the time evolution of the number of individuals for 500 independent simulations of the r-IBM presented in Section~\ref{subec.reduced.model} with parameters given at Table~\ref{table.parametres.comp} and with $d=1.8$. Amongst the 500 simulations, 492 go extinct before  time $t=895$ h. This illustrates the importance of the randomness due to the small number of individuals at the initial time. Moreover, amongst the surviving populations, we can observe randomness also in the time it takes for the population to reach a certain mass. The empirical law of the extinction time, determined from the 492 extinct populations, is given in Figure~\ref{fig.extinction.time}.

This illustrates the importance of stochastic modeling of the mutant population just after the mutation event. However, as explained in Section~\ref{links.between.fitness} and illustrated in the following section, we can use the deterministic r-PDE to determine which mutant populations can invade.

%---------------------------------------------
%---------------------------------------------

\subsection{Evolution of a one-dimensional parameter of the division kernel}
\label{sec.evol.c}

We are interested in the evolution of the division kernel, more precisely in the best  division proportion. We assume here that mutations affect only the mean proportion $c$ of the smallest daughter cell and not the minimal mass for division $\xdiv$. 
In this section, we will use the notations $c_R$ and $c_M$ instead of $\xi_R$ and $\xi_M$ for the resident and the mutant trait.
We study the invasion possibilities with respect to the resident population with trait $c_R$ and the mutant population with trait $c_M$.

\subsubsection{Pairwise invasibility plot}

\begin{figure}
\begin{center}
\includegraphics[width=6cm]{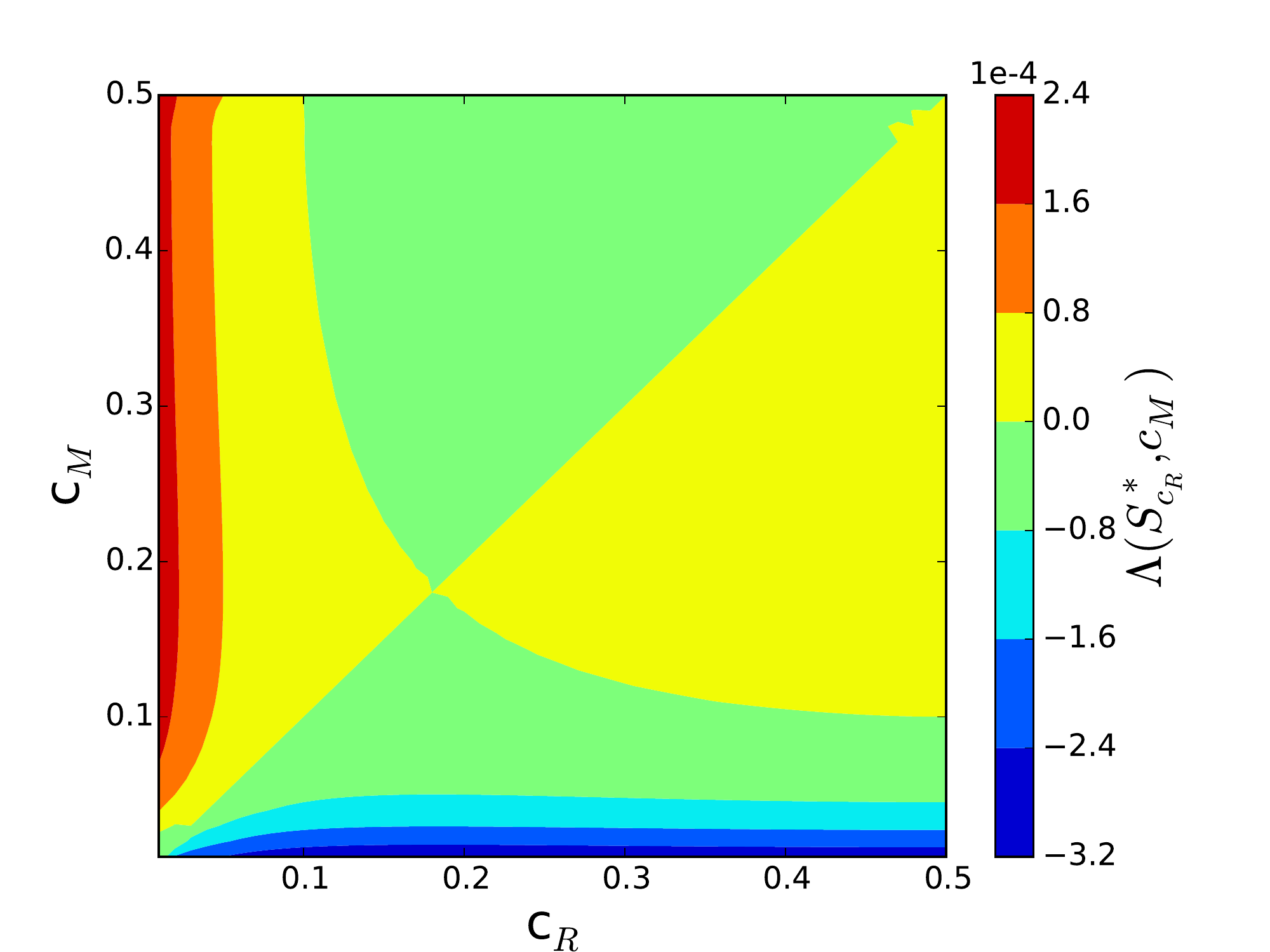}
\includegraphics[width=6cm]{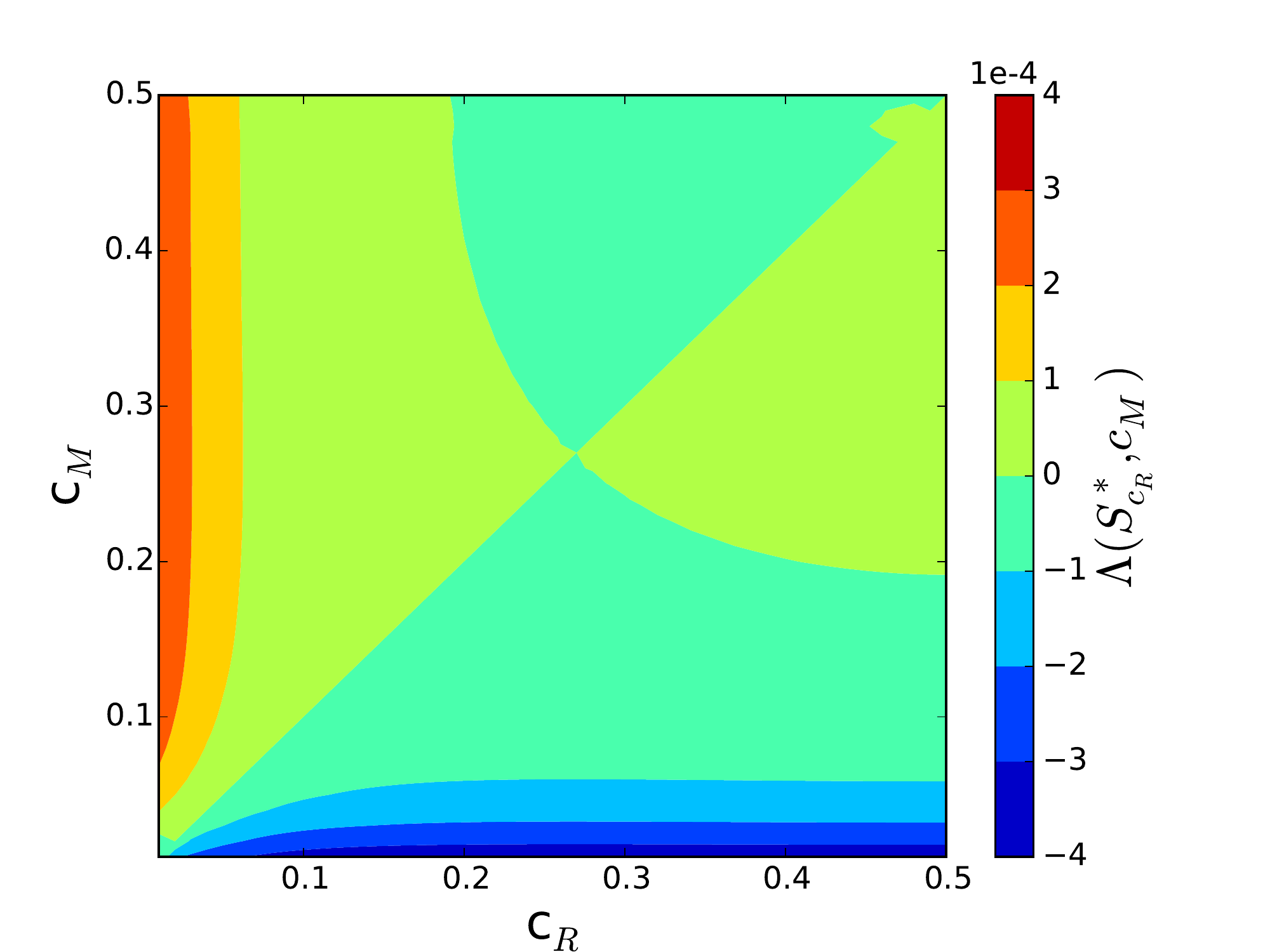}\\
\includegraphics[width=6cm]{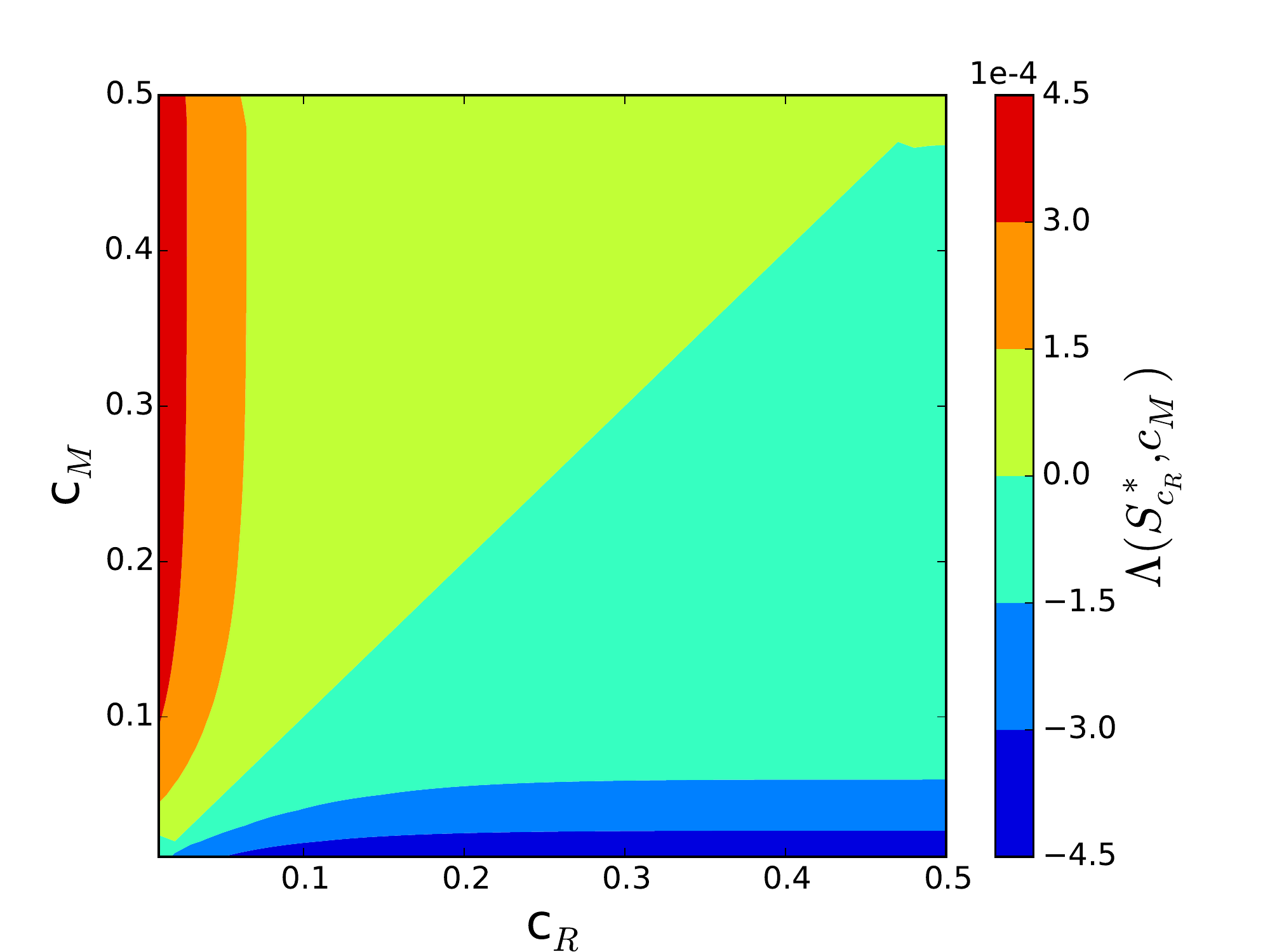}
\end{center}
\caption{Pairwise invasibility plots for the evolution parameter $c$ with parameters of Table~\ref{table.parametres.comp} and $d=1.8$ (top/left), $d=1.85$ (top/right) and $d=1.9$ (bottom).
%(Ref. of the simu: 160107160713 ; 160107160806 ; 160107160842)
}
\label{fig.PIP}
\end{figure}

Figure~\ref{fig.PIP} shows pairwise invasibility plots (PIP) for the evolutionary parameter $c$ for $d=1.8$ (top/left), $d=1.85$ (top/right) and $d=1.9$ (bottom) in the r-PDE model. It corresponds to the invasion fitness for the reduced deterministic model, i.e. the eigenvalue $\Lambda(S^*_{c_R},c_M)$ defined by~\eqref{eq.eigenproblem}, with respect to the mutant and resident traits. The regions where this eigenvalue is positive correspond to mutant and resident parameters for which the invasion of the mutant population is possible for the deterministic r-PDE model and then, as explained in Section~\ref{links.between.fitness}, also for the r-IBM. These eigenvalues do not translate into the quantitative values of the invasion probabilities. However, according to the following result \cite[Corollary 3.11]{campillo2016a}:
\begin{align}
\label{variation.inv.fit}
p^c_{S^1}(x) \leq p^c_{S^2}(x) \quad 
\Longleftrightarrow \quad r(S^1)\leq r(S^2)  \quad
\Longleftrightarrow \quad \Lambda(S^1,c)\leq \Lambda(S^2,c)\,,
\end{align}
the higher the eigenvalue is, the higher the invasion probability is.  

\bigskip

On the top-right of the plots of Figure~\ref{fig.PIP}, we observe numerical artifacts  due to the fact that the computation is numerically ill-conditioned in this part of the parameter space. Here, the difference of the global growth between the mutant and resident populations is very small.
To obtain a more accurate evaluation of the sign of the eigenvalues for this part of the parameter space, a specific numerical development would need to be undertaken, particularly in order to solve the eigenproblem in a targeted way. This exceeds the goals of the present study.

\bigskip

\begin{figure}
\begin{center}
\includegraphics[width=6cm]{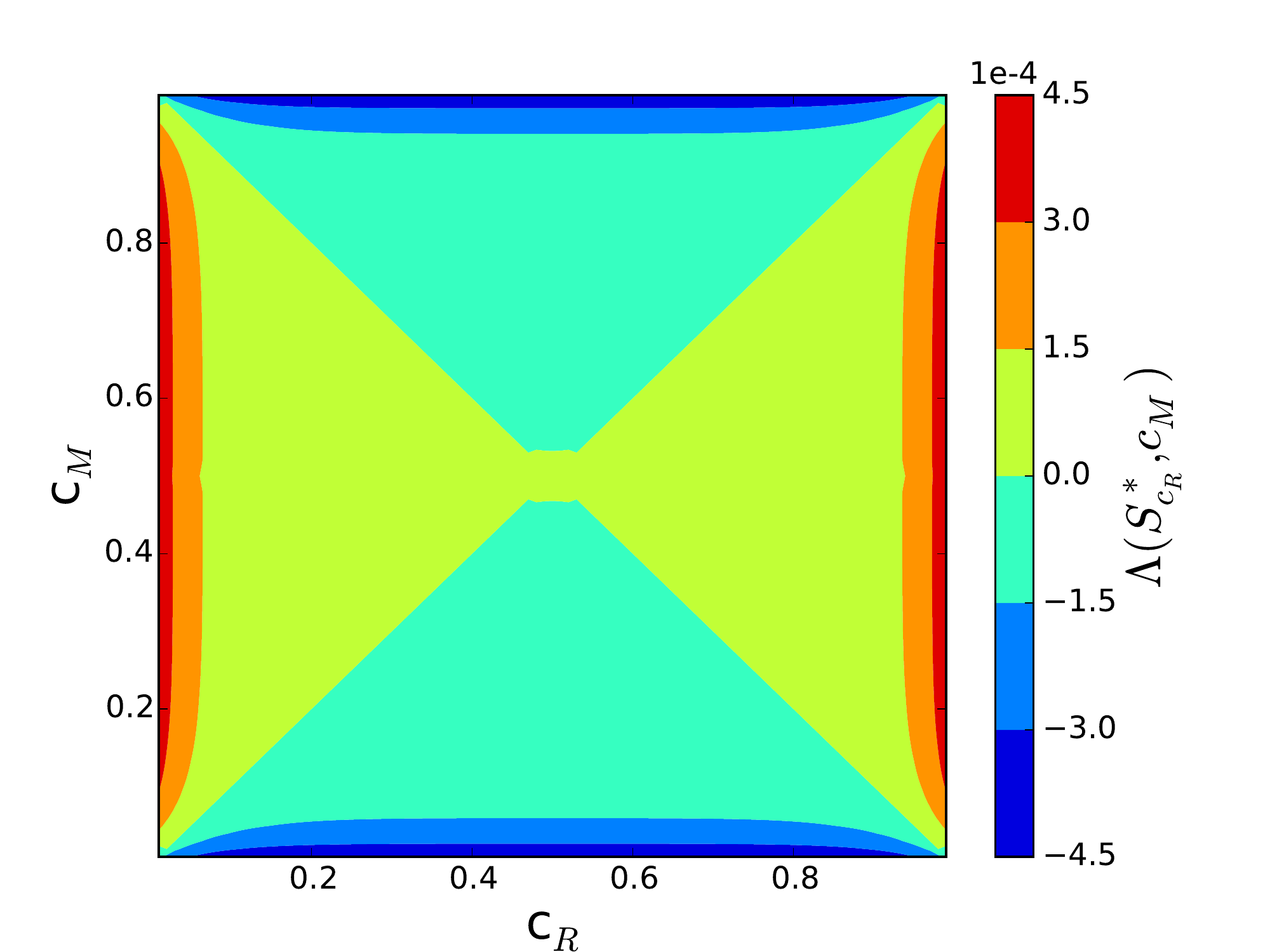}
\end{center}
\caption{Pairwise invasibility plots for the evolution parameter $c$ going from 0 to 1
 with parameters of Table~\ref{table.parametres.comp} and $d=1.9$.
\label{fig.symetry}}
\end{figure}

For each PIP of Figure~\ref{fig.PIP}, we observe the existence of one particular trait value $c^*$ which is the intersection of the main diagonal and of the second curve for which the fitness is zero. 
For the third PIP, this particular trait value is $c^*=0.5$, as illustrated in Figure~\ref{fig.symetry}, which represents, the PIP for $c$ going from 0 to 1. By symmetry of the function $q$ with respect to 0.5, this PIP is symmetric with respect to $c_R=0.5$ and $c_M=0.5$.
If we assume that the invasion fitness is locally differentiable with respect to $c_M$, then the sign of the  gradient
$$
	\textrm{Grad}(c_M) \eqdef \left[\frac{\partial \Lambda(S^*_{c_R},c_M)}{\partial c_M}\right]_{c_R=c_M}
$$
changes for $c_M=c^*$. Hence, this particular trait $c^*$  satisfies $\textrm{Grad}(c^*)=0$ and thus it is an evolutionary singular strategy (ESS) (see \cite{geritz1998a}).

Moreover, the vertical line $c_R=c^*$ is completely included in a region in which the invasion fitness $\Lambda(S^*_{c_R},c_M)$ is negative, therefore no mutant can invade the population $c^*$. Thus, the singular strategy is ESS-stable.

Furthermore, the regions above the diagonal for $c_M<c^*$ and below the diagonal for $c_M>c^*$ correspond to  positive invasion fitness. Then, if only small mutations can occur (that is $c_M$ is `close' to $c_R$), only mutants closer to $c^*$ than the resident ($c_R<c_M<c^*$ or $c^*<c_M<c_R$) can invade, and in this case the singular strategy $c^*$ is said convergence-stable.
An evolutionarily singular strategy which is both ESS-stable and convergence-stable is said to be continuously stable (see \cite{geritz1998a}).

\begin{figure}
\begin{center}
\includegraphics[width=6cm]{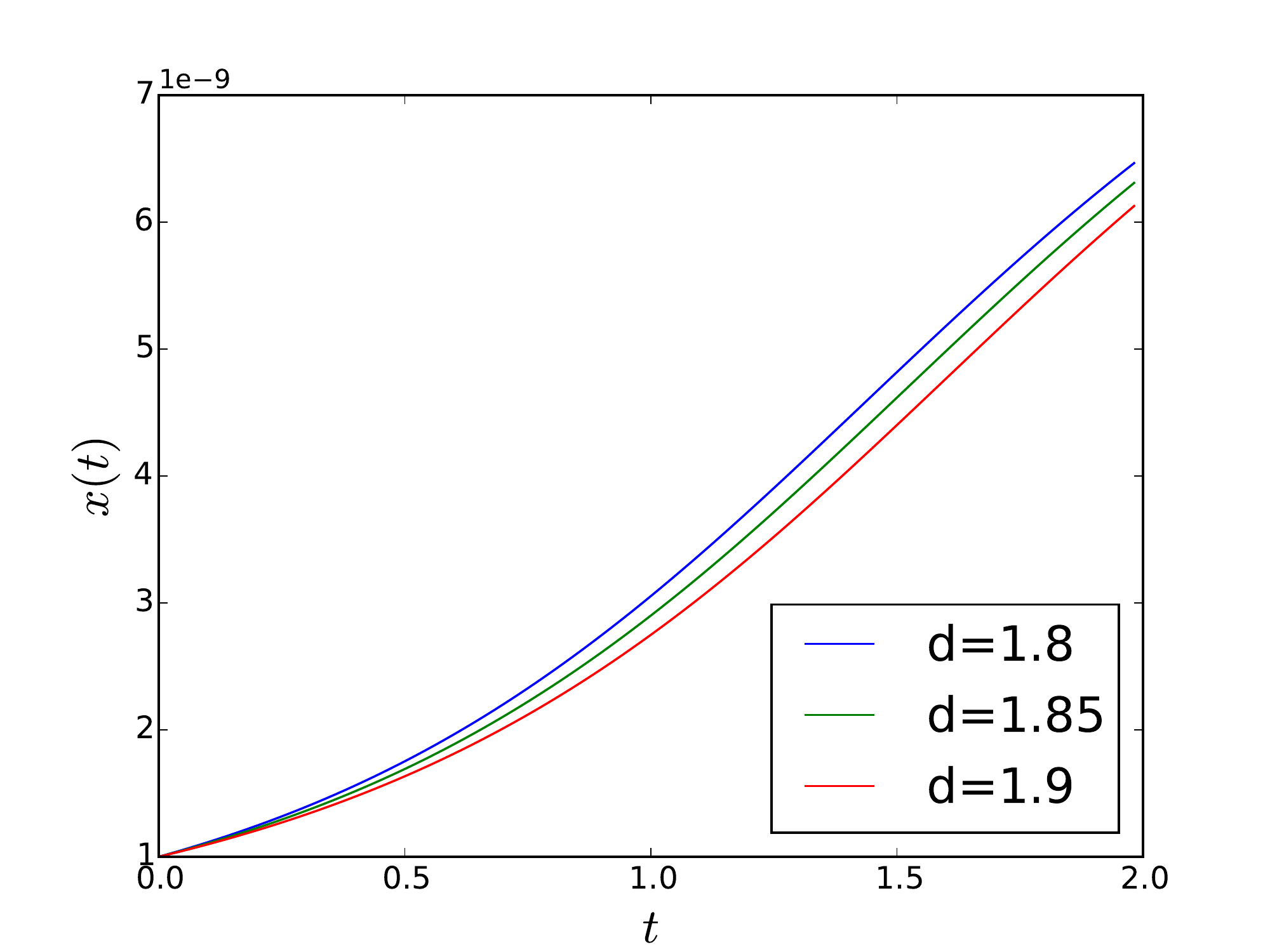}
\includegraphics[width=6cm]{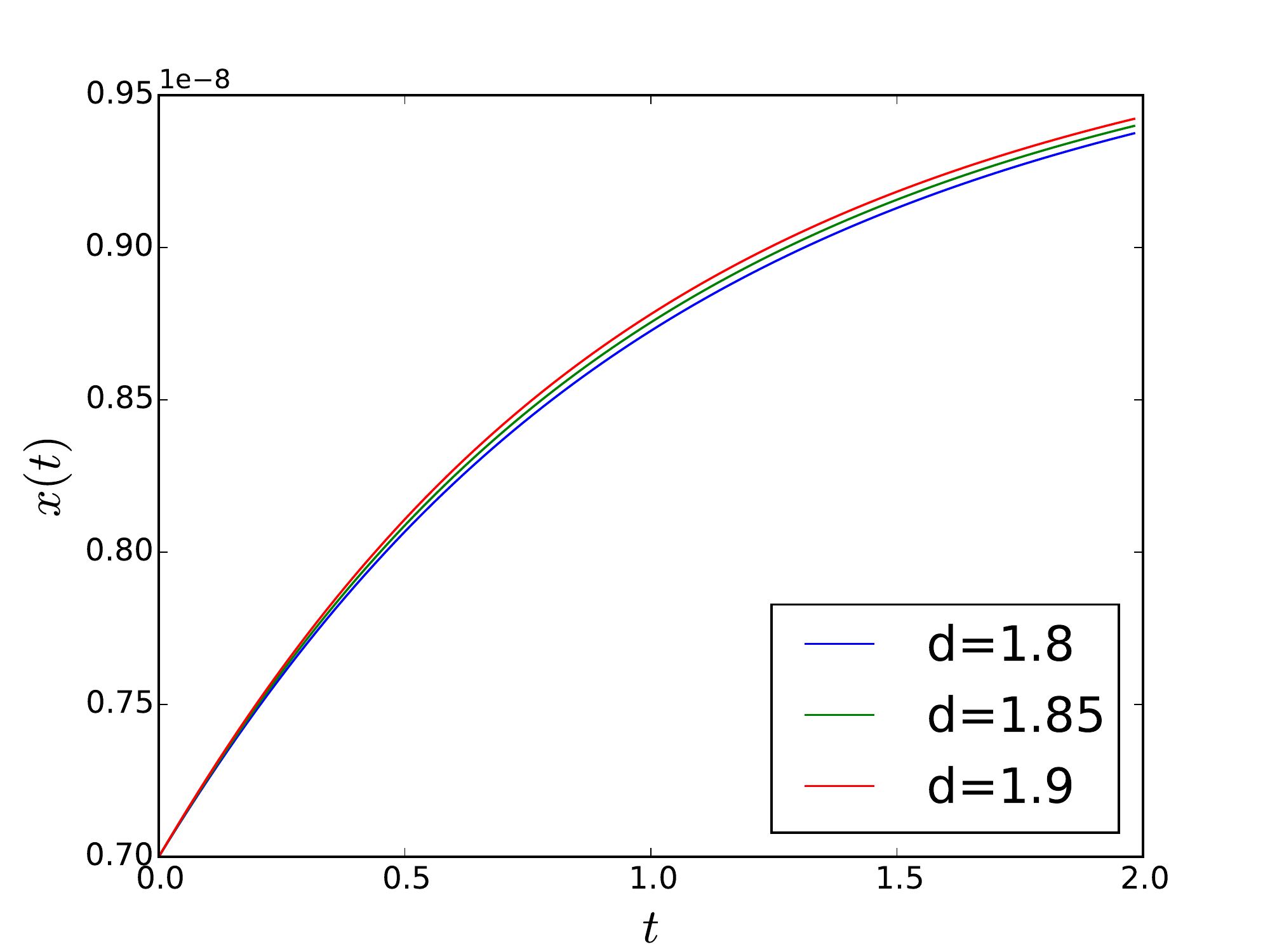}
\end{center}
\caption{Time evolution of the mass of one individual with the maximal growth (for $\rmax=1$ h$^{-1}$) $\tilde g$ given by Equation~\eqref{eq.tile.g} with $\mmax=1\times 10^{-8}$ mg, $a=1.4$ and $d=1.8$ (blue), $d=1.85$ (green) and $d=1.9$ (red) and with the initial mass $x(0)=0.1\times 10^{-8}$ mg (left) and $x(0)=0.7\times 10^{-8}$ mg (right).}
\label{fig.growthSpeed}
\end{figure}

We observe that the bigger the parameter $d$, the bigger the ESS. As illustrated in Figures~\ref{fig.tilde.g} (center) and \ref{fig.growthSpeed}, high values of $d$ favor the growth of large individuals whereas low values of $d$ favor the growth of small individuals. Then, the values of the ESS in Figure~\ref{fig.PIP} can be explained by the fact that for small values of $d$, it is better to split in a very asymmetric way  in order to produce small individuals with a relatively fast growth rate, and to maintain big individuals which will quickly divide. For big values of $d$, it is better to split in a more symmetrical way in order to produce medium sized individuals with both a fast growth rate and a fast splitting rate. 

PIP's are antisymmetric with respect to the first diagonal, meaning that mutual invasibility is not possible, and also mutual non-invasibility is not possible, which hints at optimising evolution \citep{geritz1998a, metz2008b}.

%---------------------------------------------
%---------------------------------------------

%---------------------------------------------
%---------------------------------------------

%---------------------------------------------
%---------------------------------------------

\subsubsection{State at equilibrium and ESS}
\label{subsubsec.eq}

\begin{figure}
\begin{center}
\includegraphics[width=6cm]{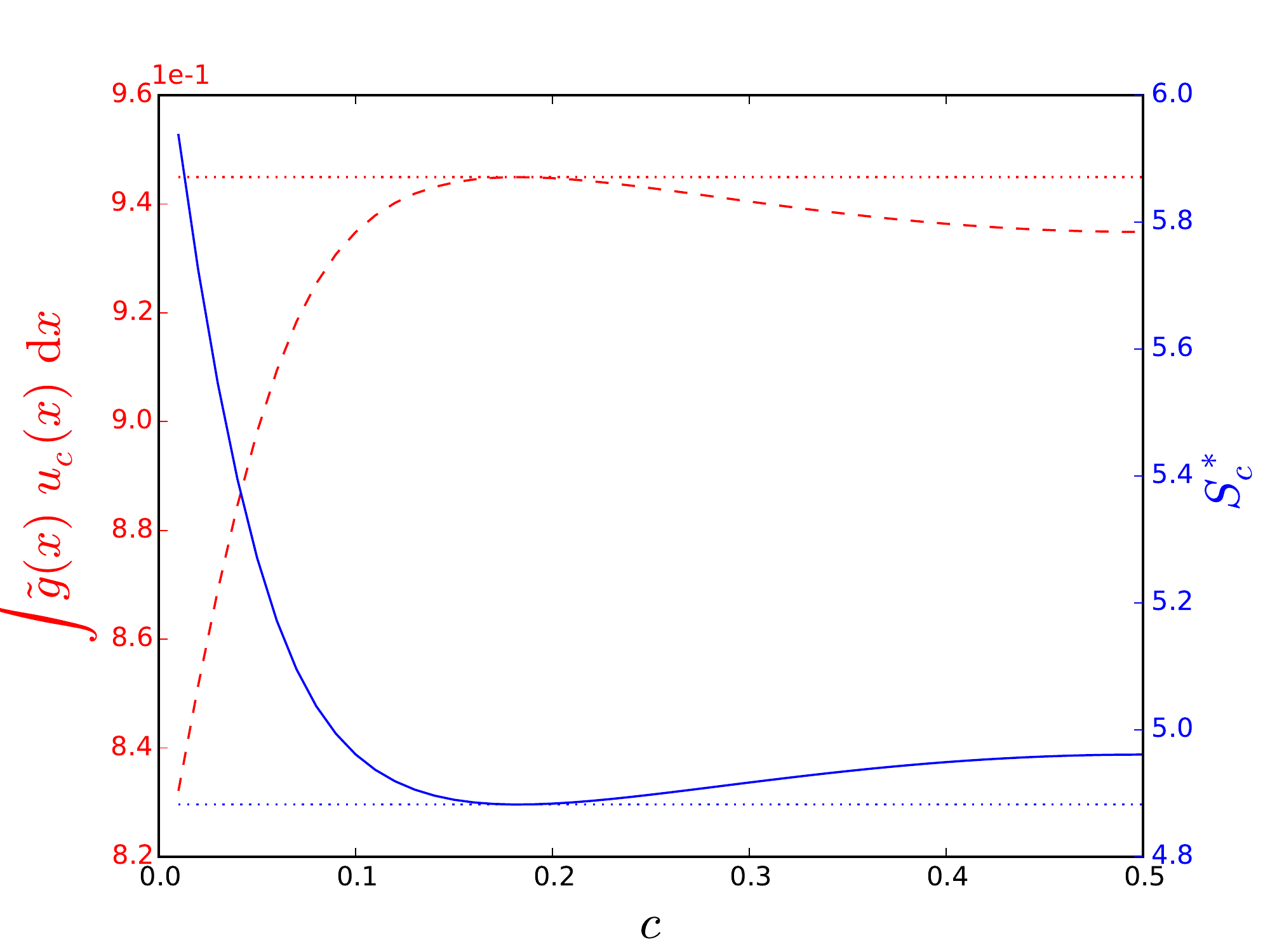}\\
\includegraphics[width=6cm]{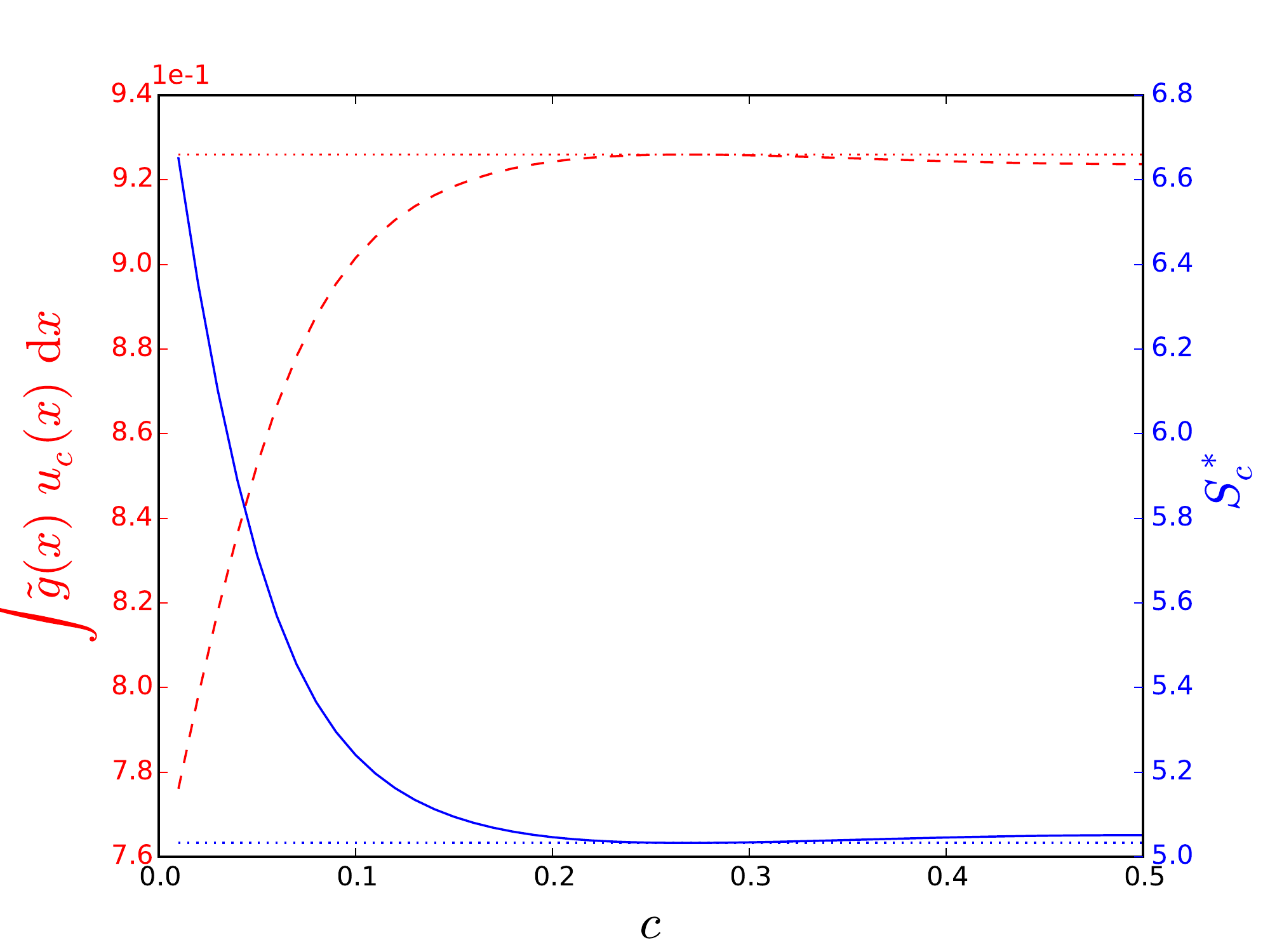}
\includegraphics[width=6cm]{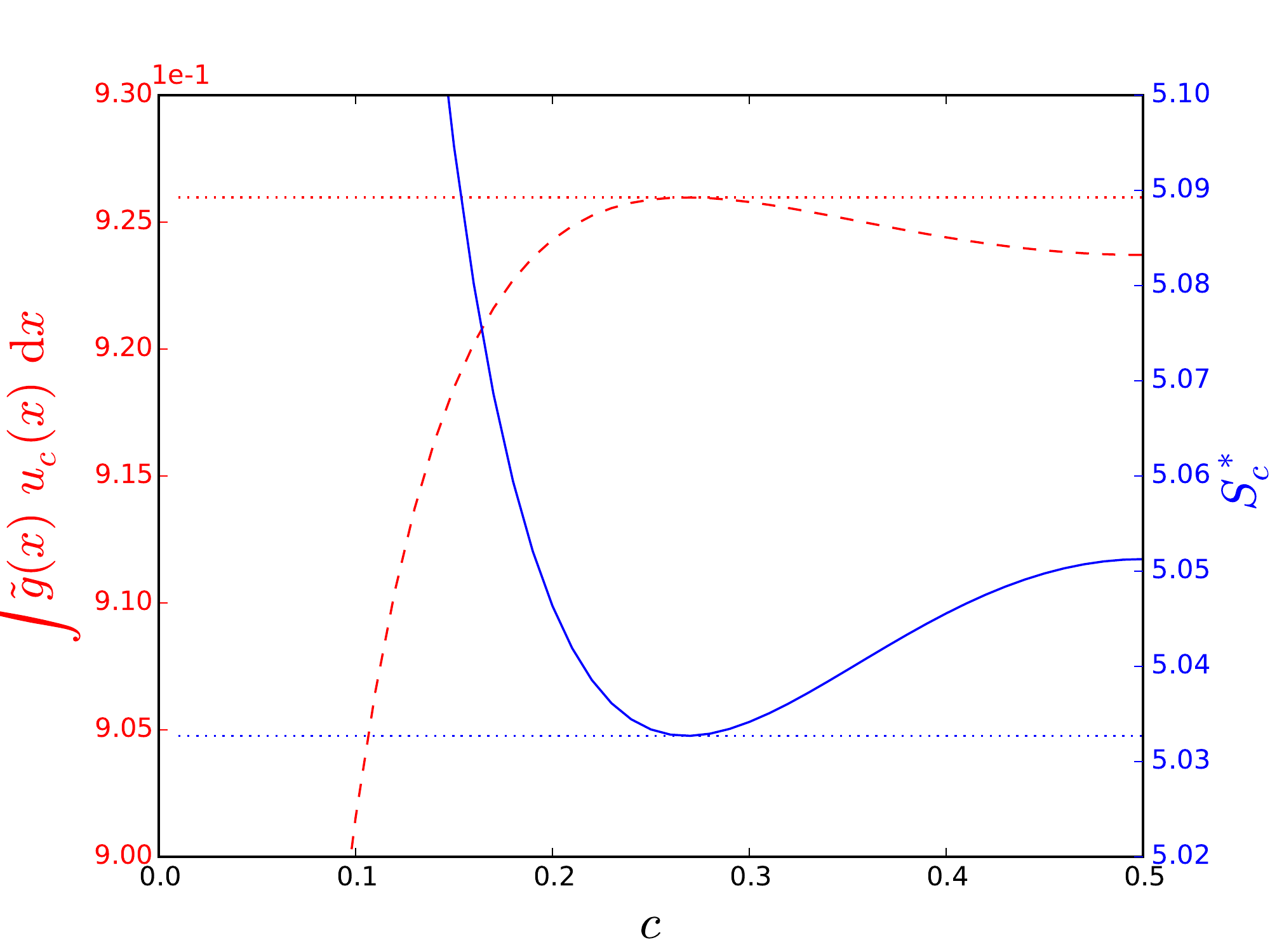}\\
\includegraphics[width=6cm]{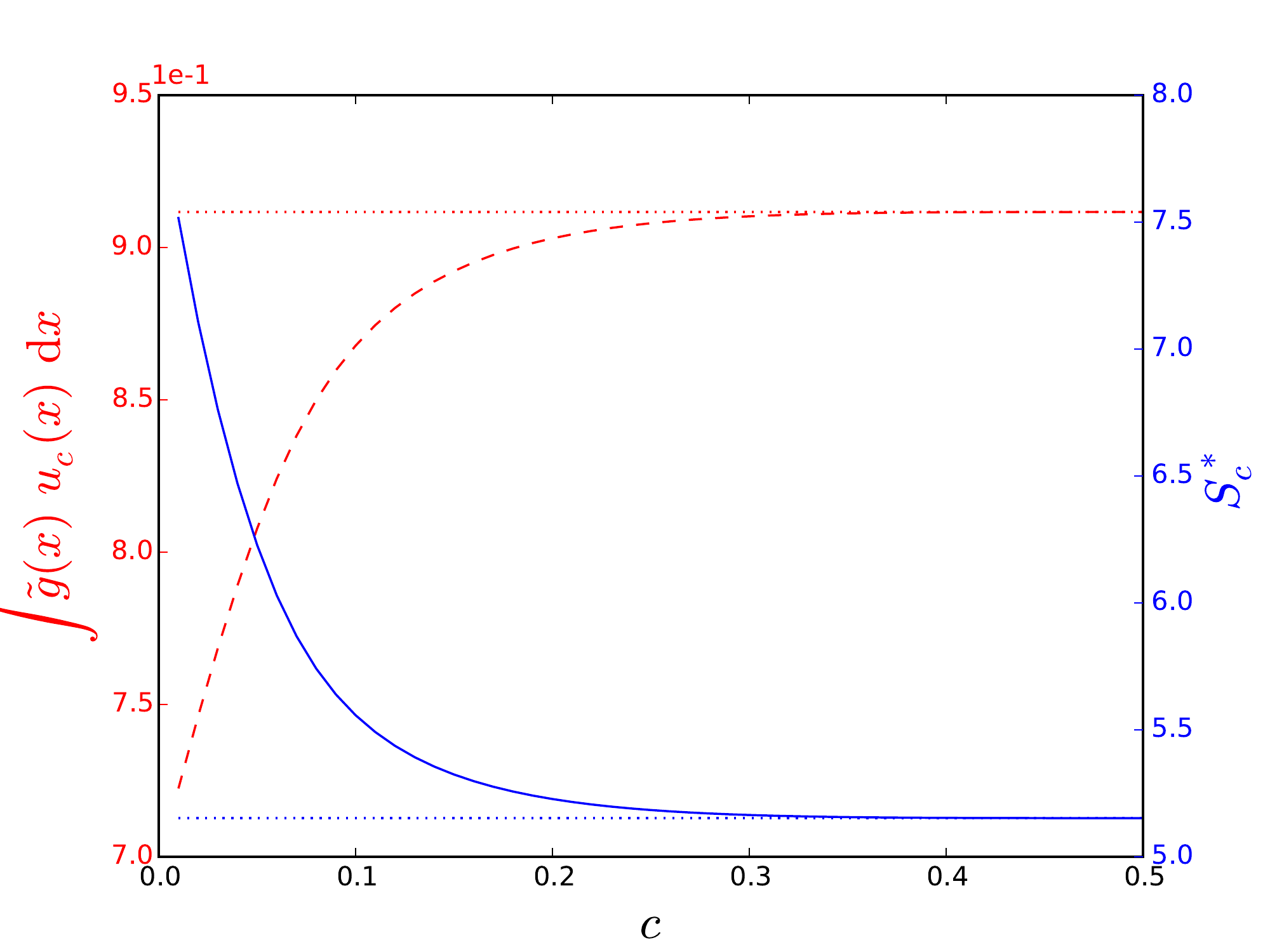}
\includegraphics[width=6cm]{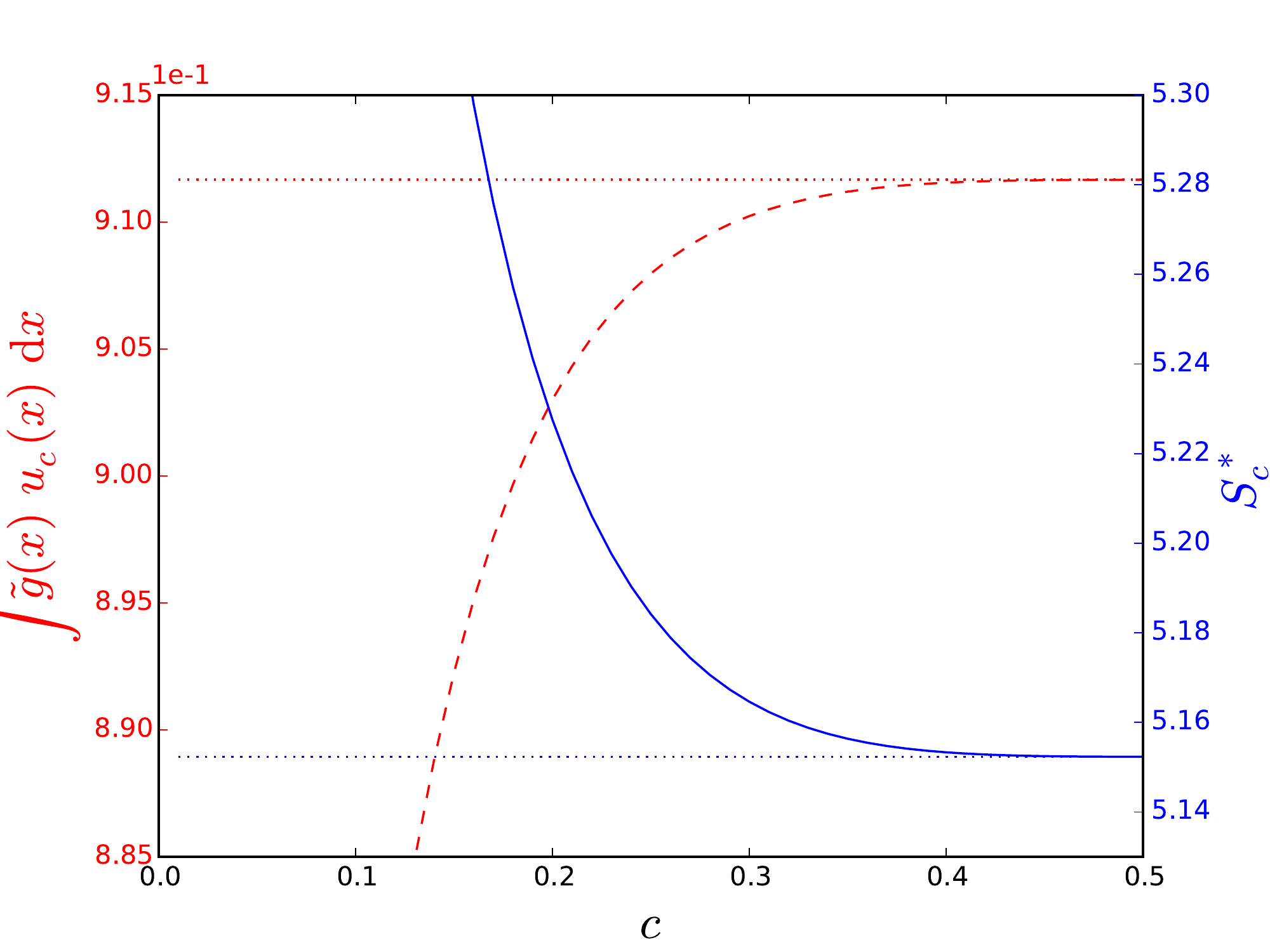}
\end{center}
\caption{Substrate concentration at the stationary state (continuous blue curves) and $\int_0^{\mmax}\tilde g(x)\,u_c(x)\,\dif x $ (dashed red curves) where $u_c$ is defined by \eqref{def_uc} with respect to the division proportion parameter $c$ for $d=1.8$ (top), $d=1.85$ (middle) and $d=1.9$ (bottom). The blue and red dotted lines respectively correspond to the minimum and the maximum of these curves. Right plots are zooms on the left plots in order to observe the minimum and maximum of curves.}
\label{fig.s_star}
\end{figure}

The result stated in~\eqref{variation.inv.fit} allows one to only compute the substrate concentrations at the equilibrium $S^*_{c_R}$ and $S^*_{c_M}$ in order to determine if the mutant population $c_M$ can invade the resident one $c_R$. Indeed, as $\Lambda(S^*_{c_M},c_M)=0$ and as $S\mapsto r(S)$ defined by \eqref{monod} is increasing, we know that $\Lambda(S^*_{c_R},c_M)\geq 0$ if and only if $S^*_{c_R} \geq S^*_{c_M}$.
Hence, to study the invasion feasibility by using the sign of the PIP, it is sufficient to compute the substrate concentration at equilibrium with respect to the evolution parameter $c$. Figure~\ref{fig.s_star} represents the substrate concentration at equilibrium (continuous blue curves) with respect to the trait $c$ for the three values of $d$ previously considered.
For each $c_R$, the invasion fitness $\Lambda(S^*_{c_R},c_M)$ will be positive for each $c_M$ such that $S^*_{c_M} < S^*_{c_R}$.
Moreover, the ESS corresponds to the trait in which the minimum of the substrate concentration at the equilibrium $S^*_c$ is reached. The fact that a minimum is reached for the ESS also allows one to characterize it as convergence-stable.

%--------------------------------------------
%---------------------------------------------

\bigskip

The normalized density weighted by the mass at the equilibrium for the r-PDE is defined by the function $x\mapsto x\,u_c(x)$ with
\begin{align}
\label{def_uc}
	u_c(x) = \frac{r^*_c(x)}{\int_0^{\mmax} z\,r^*_c(z)\,\dif z} 
\end{align}
where $r^*_c$ is the population density at the equilibrium in the chemostat, that is $(S^*_c,r^*_c)$ is solution of
\begin{align} 
\label{eq.eid.substrat_eq}
	&
	D\,(\Sin-S^*_c)-\frac kV \int_0^{\mmax} g(S^*_c,x)\,
	r^*_c(x)\,\dif x=0\,,
\\
\label{eq.eid.pop_eq}
	&
	\frac{\partial}{\partial x} \bigl( g(S^*_c,x)\,r^*_c(x)\bigr)
	+ \bigl(\taudiv(\xdiv,x)+D \bigr)\,r^*_c(x)
	=  
	2\,\int_x^{\mmax}
		\frac{\taudiv(\xdiv,z)}{z} \, 
		q_c\left(\frac{x}{z} \right) \,
		r^*_c(z)\,\dif z \,.
\end{align}
Then, by integrating Equation~\eqref{eq.eid.pop_eq} with respect to $x\,\dif x$, we get
\begin{align*} 
	\int_0^{\mmax} g(S^*_c,x)\,r^*_c(x)\,\dif x
	=  
	D\,\int_0^{\mmax} x\,r^*_c(x)\,\dif x\,
\end{align*}
then, by \eqref{def_uc},
\begin{align*}
r(S^*_c)\,\int_0^{\mmax}\tilde g(x)\,u_c(x)\,\dif x 
= D \,.
\end{align*}
Therefore, the parameter $c$ which minimizes the function $c \mapsto r(S^*_c)$, is also the one which maximizes the function $c\mapsto \int_0^{\mmax}\tilde g(x)\,u_c(x)\,\dif x $, as illustrated in Figure~\ref{fig.s_star}.
Hence, as the function $r$ is increasing, minimizing the substrate concentration $S_c^*$ is equivalent to maximizing the bacterial growth.

%%%%%%%%%%%%%%%%%%%%%%%%%%%%%%%%%%%%%%%%%%%%%%%%%%%%%%%%%%%%%%%%%%%%%%
%%%%%%%%%%%%%%%%%%%%%%%%%%%%%%%%%%%%%%%%%%%%%%%%%%%%%%%%%%%%%%%%%%%%%%
\subsection{Evolution of a two-dimensional parameter}
\label{sec.evol.c.y}
%%%%%%%%%%%%%%%%%%%%%%%%%%%%%%%%%%%%%%%%%%%%%%%%%%%%%%%%%%%%%%%%%%%%%%
%%%%%%%%%%%%%%%%%%%%%%%%%%%%%%%%%%%%%%%%%%%%%%%%%%%%%%%%%%%%%%%%%%%%%%

The method of Section~\ref{subsubsec.eq} extends straightforwardly to the evolutionary analysis of multi-dimensional traits.
Traits for which the substrate concentration reaches a (global) minimum correspond to a (global) convergence-stable singular strategy and the traits for which the substrate concentration reaches a (global) maximum correspond to a (global) repulsive singular strategy in the sense that the trait will tend to evolve in a way to draw away from this trait.

\medskip

\begin{figure}
\begin{center}
\includegraphics[width=6.5cm]{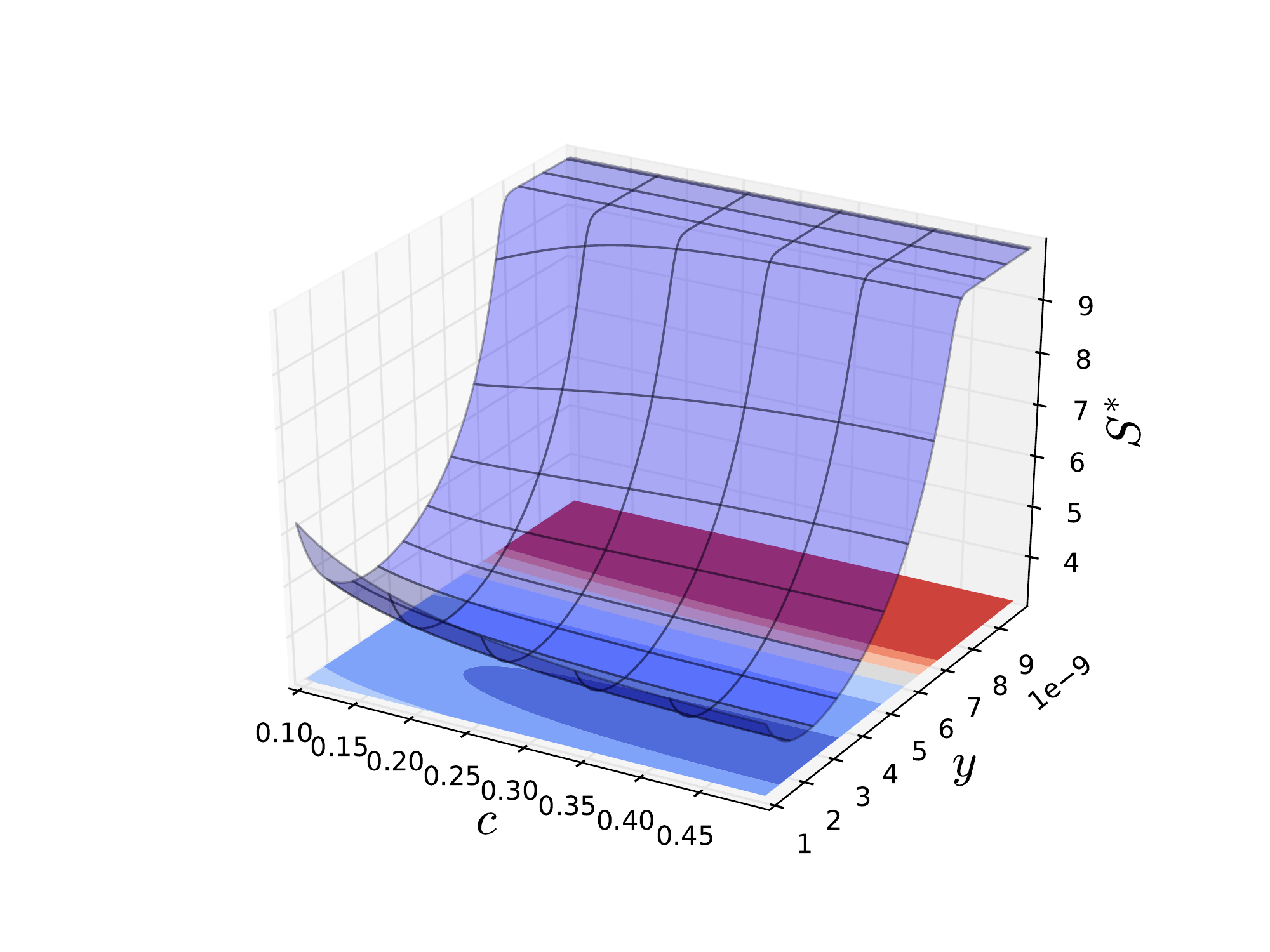}
\includegraphics[width=5.5cm]{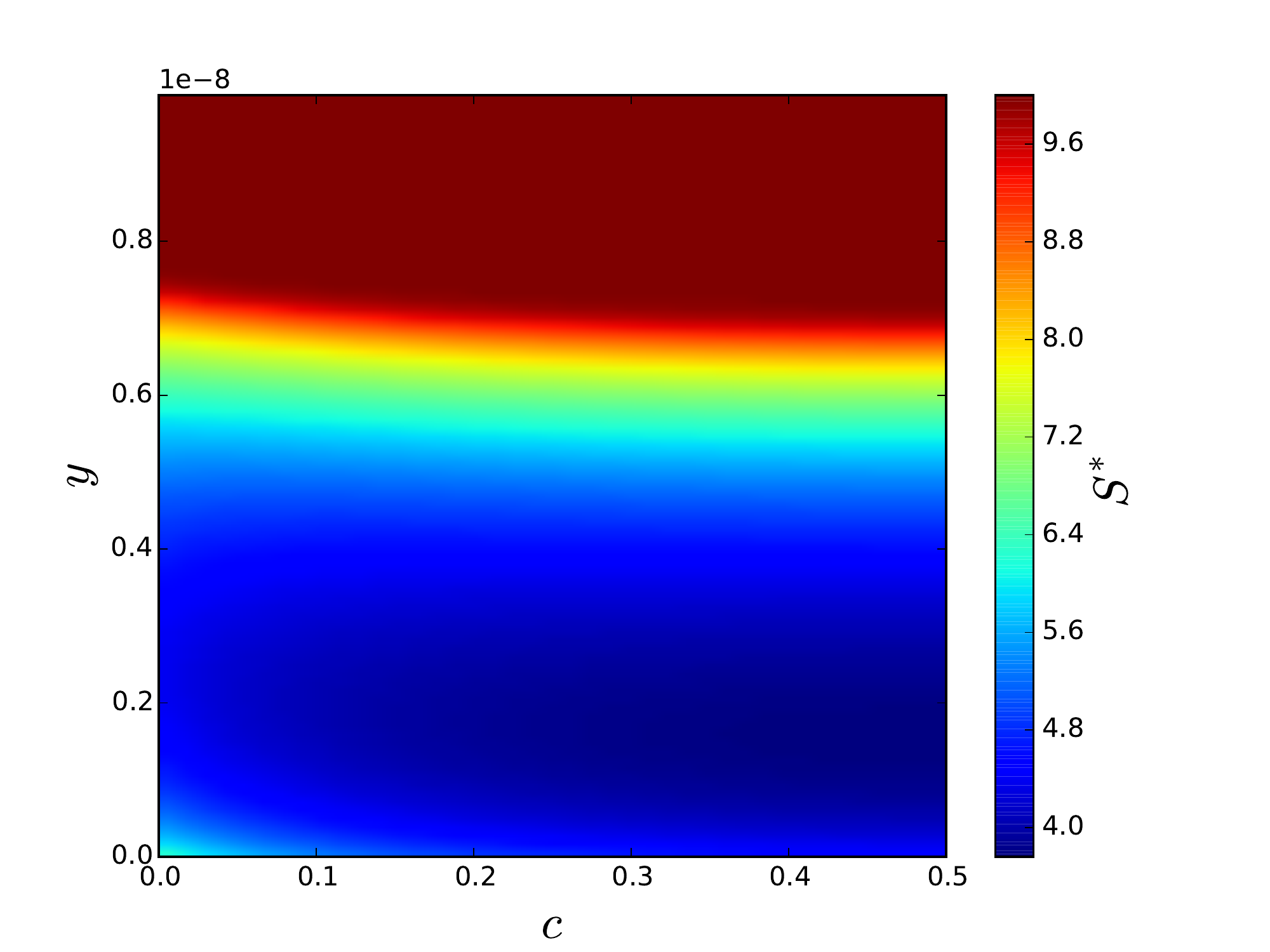}
\end{center}
\caption{Substrate concentration at the equilibrium with respect to the mean proportion $c$ and the minimal mass for division $y$. Simulation is performed with parameters of Table~\ref{table.parametres.comp} and with $d=1.8$.}
\label{fig.sub.2D}
\end{figure}

Here, we give an illustration for the two-dimensional evolution parameter $(c,\xdiv)$, that is, the mutations affect not only the mean proportion of the smallest daughter cell during the division but also the minimal mass for which a division is possible. Figure~\ref{fig.sub.2D} represents the substrate concentration at equilibrium, i.e. the stationary state of Equation~\eqref{eq.eid.substrat} of the system~\eqref{eq.eid.substrat}-\eqref{eq.eid.pop}, with respect to $c$ and $\xdiv$. The minimum of this function is reached in $(c,\xdiv)^*=(0.5,0.24\times 10^{-8})$, which is a convergence-stable ESS. 
Moreover, the division rate is sufficiently large for mass larger than $\xdiv$ ($\bar b=20$ h$^{-1}$);  bacteria divide very quickly when they reach mass $\xdiv$.
Hence, the optimal behavior of the population for this two-dimensional evolution parameter is to split in a (noisy) symmetrical way when the bacterial reaches the size $x=0.24\times 10^{-8}$ g.

%---------------------------------------------
%---------------------------------------------

%%%%%%%%%%%%%%%%%%%%%%%%%%%%%%%%%%%%%%%%%%%%%%%%%%%%%%%%%%%%%%%%%%%%%%
%%%%%%%%%%%%%%%%%%%%%%%%%%%%%%%%%%%%%%%%%%%%%%%%%%%%%%%%%%%%%%%%%%%%%%
\section{Discussion}
\label{ccl}
%%%%%%%%%%%%%%%%%%%%%%%%%%%%%%%%%%%%%%%%%%%%%%%%%%%%%%%%%%%%%%%%%%%%%%
%%%%%%%%%%%%%%%%%%%%%%%%%%%%%%%%%%%%%%%%%%%%%%%%%%%%%%%%%%%%%%%%%%%%%%

In this article, we have proposed an approach to study mutations affecting the cell division in a limited and controlled environment such as the chemostat. Our objective was to determine in which circumstances a mutant population could replace a resident one. The main difficulty is to determine the suitable modeling approach to describe the problem. Stochastic representations are required to model a population in small size, which is the case at the very moment of the appearance of a mutant strain. In contrast with deterministic representations, stochastic representations allow for an explicit derivation of the extinction probability of the mutant population at its appearance.
However, stochastic representations are less straightforward to compute and analyze than their deterministic counterparts. Moreover, deterministic models, generally understood in a biological context as an asymptotic limit in large population size of a stochastic model, are essential objects in order to apprehend the stochastic models.

We have proposed a model reduction technique that takes advantage of both stochastic and deterministic approaches: with the help of a previously published mathematical result, one can characterize the mutant strains that can invade a resident population according to the stochastic model by simulating the deterministic model. 

Using the previously described reduced model, we have represented the mutual invasion possibilities with respect to the mutant and the resident traits through a standard pairwise invasibility plot (PIP). This plot has permitted us to determine an evolutionary singular strategy (ESS). This strategy achieves the best substrate degradation, that is the minimal substrate concentration in the chemostat at the stationary state obtained by adjusting the evolutionary trait. This trait will then characterize the “best” population growth. Following the argument of substrate concentration at the stationary state and using a mathematical result previously proved in \cite{campillo2016a},
 we have proposed a more efficient simulation method to determine the invasion possibilities. In fact, we have drawn the PIP by computing, for each resident and mutant trait, first, the stationary state of the resident population and second, the growth rate of the mutant population in this stationary state. However, as illustrated in this article, it is sufficient to compare the substrate concentrations at the stationary state for both traits in order to determine if the mutant population is able to invade or not. Nevertheless, information is lost with this method as we can obtain qualitative results, such as the “shape” of the PIP, but no quantitative results, such as the value of the growth rate generally represented in this kind of plot.

We also loose information on the stochastic model, and more precisely on the survival probability, when we use the deterministic model to represent the PIP. A challenge for future work is to establish a relationship between the survival probability and the eigenvalue as for the birth-death process for which the survival probability is equal to $\left[\Lambda/b \right]_+$ where $\Lambda = b - d$ where $b$ and $d$ are the birth and death rates.

We now comment some assumptions, which are typically made in adaptive dynamics and which we made in this study.
First that the resident population is large and second, that the mutations are relatively rare. Although these assumptions may be debatable, however, they allow one to determine the convergence-stable ESS, which corresponds to the optimal trait of the population independently of these assumptions. Moreover, they allow one to compare two different traits in order to determine which one is best able to degrade the substrate, while remaining independent of both assumptions.

One other assumption made in this approach is that the deterministic equilibrium of the resident population is non trivial, i.e. that the population does not go extinct. It could be interesting to extend the study to more general models, for example for a non monotonic
Haldane growth rates $r(S)\eqdef r_{\max} \, \frac{S}{K_r + S+\alpha/S^2}$, for which a mutant population can invade the chemostat but not starting from any initial conditions.

The reduction technique proposed here on a chemostat example has the advantage to be generic and can be applied to other bacterial ecosystems or other population cell models. In additions to methods used in standard adaptive dynamics, this technique is based on the characterization of the deterministic model as the large population size limit of the stochastic model and on the link between the invasion criteria for the deterministic models and the corresponding definition for the stochastic models.
The first point is related to technique of deriving a partial differential equation as a limit of the stochastic individual based model \citep{fournier2004a, champagnat2006a, tran2008a}. 
The mathematical approach used in our previous article \citep{campillo2015a} to link the two invasion criteria is relatively general and can be applied to models in which there is no interaction between individuals. This is, in general, the case for models in adaptive dynamics for which, in the initial phase of mutant invasion, the interactions between mutants can be ignored, and the resident population can be viewed to provide a constant environment for the mutant.

%%%%%%%%%%%%%%%%%%%%%%%%%%%%%%%%%%%%%%%%%%%%%%%%%%%%%%%%%%%%%%%%%%%%%%
%%%%%%%%%%%%%%%%%%%%%%%%%%%%%%%%%%%%%%%%%%%%%%%%%%%%%%%%%%%%%%%%%%%%%%
\section*{Acknowledgements}
%%%%%%%%%%%%%%%%%%%%%%%%%%%%%%%%%%%%%%%%%%%%%%%%%%%%%%%%%%%%%%%%%%%%%%
%%%%%%%%%%%%%%%%%%%%%%%%%%%%%%%%%%%%%%%%%%%%%%%%%%%%%%%%%%%%%%%%%%%%%%
The work of Coralie Fritsch was partially supported by the Meta-omics of Microbial Ecosystems (MEM) metaprogram of INRA and by the Chair ``Mod\'elisation Math\'ematique et Biodiversit\'e'' of VEOLIA Environment, \'Ecole Polytechnique, Mus\'eum National d'Histoire Naturelle and Fondation X.
The study was supported financially by the Academy of Finland (Grant no. 250444 to Otso Ovaskainen) and the Research Council of Norway (CoE grant no. 223257).

\bibliographystyle{apalike}
%\bibliography{lib/coralie}

\begin{thebibliography}{}

\end{thebibliography}


\begin{thebibliography}{}

\bibitem[Campillo et~al., 2016a]{campillo2015a}
Campillo, F., Champagnat, N. and Fritsch, C. (2016a).
\newblock Links between deterministic and stochastic approaches for invasion in
  growth-fragmentation-death models.
\newblock \textit{Journal of Mathematical Biology}.
\newblock , 73(6):1781--1821.

\bibitem[Campillo et~al., 2016b]{campillo2016a}
Campillo, F., Champagnat, N. and Fritsch, C. (2016b).
\newblock On the variations of the principal eigenvalue and the probability of
  survival with respect to a parameter in growth-fragmentation-death models.
\newblock \textit{ArXiv Mathematics e-prints}.
\newblock arXiv/1601.02516 [math.AP].

\bibitem[Campillo and Fritsch, 2015]{campillo2015b}
Campillo, F. and Fritsch, C. (2015).
\newblock Weak convergence of a mass-structured individual-based model.
\newblock \textit{Applied Mathematics \& Optimization}, 72(1):37--73.

\bibitem[{Champagnat, 2006}]{champagnat2006a}
Champagnat, N. (2006).
\newblock {A microscopic interpretation for adaptive dynamics trait substitution sequence models}.
\newblock \textit{Stochastic Processes and their Applications}, 116:1127--1160.
  
\bibitem[Champagnat et~al., 2014]{champagnat2014a}
Champagnat, N., Jabin, P.-E. and M{\'e}l{\'e}ard, S. (2014).
\newblock Adaptation in a stochastic multi-resources chemostat model.
\newblock \textit{Journal de Math{\'e}matiques Pures et Appliqu{\'e}es},
  101(6):755--788.

\bibitem[Dieckmann and Law, 1996]{dieckmann1996a}
Dieckmann, U. and Law, R. (1996).
\newblock The dynamical theory of coevolution: a derivation from stochastic
  ecological processes.
\newblock \textit{Journal of Mathematical Biology}, 34(5):579--612.

\bibitem[Doebeli, 2002]{Doebli2002}
Doebeli, M. (2002).
\newblock A model for the evolutionary dynamics of cross-feeding polymorphisms
  in microorganisms.
\newblock \textit{Population Ecology}, 44(2):59--70.

\bibitem[Doumic, 2007]{Doumic2007a}
Doumic, M. (2007).
\newblock Analysis of a population model structured by the cells molecular
  content.
\newblock \textit{Mathematical Modelling of Natural Phenomena}, 2:121--152.

\bibitem[Doumic~Jauffret and Gabriel, 2010]{Doumic2010a}
Doumic~Jauffret, M. and Gabriel, P. (2010).
\newblock Eigenelements of a general aggregation-fragmentation model.
\newblock \textit{Mathematical Models and Methods in Applied Sciences},
  20(05):757--783.

\bibitem[{Fournier and M\'el\'eard, 2004}]{fournier2004a}
Fournier, N. and M\'el\'eard, S. (2004).
\newblock {A microscopic probabilistic description of a locally regulated population and macroscopic approximations}.
\newblock \textit{Annals of Applied Probability}, 14:1880--1919.

\bibitem[Fredrickson et~al., 1967]{Fredrickson1967a}
Fredrickson, A.~G., Ramkrishna, D. and Tsuchiya, H.~M. (1967).
\newblock Statistics and dynamics of procaryotic cell populations.
\newblock \textit{Mathematical Biosciences}, 1(3):327--374.

\bibitem[Fritsch, 2014]{fritschThesis}
Fritsch, C. (2014).
\newblock \textit{Approches probabilistes et num\'eriques de mod\`eles
  individus-centr\'es du chemostat}.
\newblock Th\`ese, {Universit{\'e} Montpellier 2}.

\bibitem[Fritsch et~al., 2015]{fritsch2015a}
Fritsch, C., Harmand, J. and Campillo, F. (2015).
\newblock A modeling approach of the chemostat.
\newblock \textit{Ecological Modelling}, 299:1--13.

\bibitem[Geritz et~al., 1998]{geritz1998a}
Geritz, S. A.~H., Ksidi, E., Mesz{\'e}na, G. and Metz, J. A.~J. (1998).
\newblock Evolutionarily singular strategies and the adaptive growth and
  branching of the evolutionary tree.
\newblock \textit{Evolutionary Ecology}, 12:35--57.

\bibitem[Metz, 2008]{metz2008a}
Metz, J. (2008).
\newblock Fitness.
\newblock In Fath, S. E. J.~D., editor, \textit{Encyclopedia of Ecology}, pages
  1599--1612. Academic Press, Oxford.

\bibitem[Metz et~al., 1996]{Metz1996a}
Metz, J. A.~J., Geritz, S. A.~H., Mesz{\'e}na, G., Jacobs, F. J.~A. and van
  Heerwaarden, J.~S. (1996).
\newblock Adaptive dynamics: A geometric study of the consequences of nearly
  faithful reproduction.
\newblock In van Strien, S.~J. and Verduyn-Lunel, S.~M., editors, \textit{
  Stochastic and spatial structures of dynamical systems (Amsterdam, 1995)},
  pages 183--231. North-Holland.

\bibitem[{Metz et~al., 2008}]{metz2008b}
Metz, J.A.J., Mylius, R.M. and Diekmann, O. (2008).
\newblock When does evolution optimize?
\newblock \textit{Evolutionary Ecology Research}, 10:629--654.
  
\bibitem[Metz et~al., 1992]{metz1992a}
Metz, J.~A., Nisbet, R.~M. and Geritz, S.~A. (1992).
\newblock How should we define `fitness' for general ecological scenarios?
\newblock \textit{Trends in Ecology \&\ Evolution}, 7(6):198--202.
  
\bibitem[Michel, 2005]{Michel2005a}
Michel, P. (2005).
\newblock Fitness optimization in a cell division model.
\newblock \textit{Comptes Rendus de l'Académie des Sciences de Paris, S\'erie I},
  341(12):731--736.

\bibitem[Michel, 2006]{michel2006a}
Michel, P. (2006).
\newblock Optimal proliferation rate in a cell division model.
\newblock \textit{Mathematical Modelling of Natural Phenomena}, 1:23--44.

\bibitem[Mirrahimi et~al., 2012]{Mirrahimi2012}
Mirrahimi, S., Perthame, B. and Wakano, J. (2012).
\newblock Evolution of species trait through resource competition.
\newblock \textit{Journal of Mathematical Biology}, 64(7):1189--1223.

\bibitem[Monod, 1950]{Monod1950a}
Monod, J. (1950).
\newblock La technique de culture continue, th{\'e}orie et applications.
\newblock \textit{Annales de l'Institut Pasteur}, 79(4):390--410.

\bibitem[Novick and Szilard, 1950]{Novick1950a}
Novick, A. and Szilard, L. (1950).
\newblock Description of the chemostat.
\newblock \textit{Science}, 112(2920):715--716.

\bibitem[Perthame, 2007]{perthame2007a}
Perthame, B. (2007).
\newblock \textit{Transport Equations in Biology}.
\newblock Birkh{\"a}user.

\bibitem[Ramkrishna, 1979]{Ramkrishna1979a}
Ramkrishna, D. (1979).
\newblock Statistical models of cell populations.
\newblock In Ghose, T., Fiechter, A., and Blakebrough, N., editors, \textit{
  Advances in Biochemical Engineering}, volume~11, pages 1--47.
  Springer-Verlag, Berlin Heidelberg New York.

\bibitem[{Tran, 2008}]{tran2008a}
Tran, V.C. (2008).
\newblock {Large population limit and time behavior of a stochastic particle model describing an age-structured population}.
\newblock \textit{ESAIM}, 12:345--386.
  
\end{thebibliography}

\end{document}